\documentclass[aps,prb,twocolumn,amsmath,amssymb,groupedaddress]{revtex4}
\usepackage{graphics,epsfig}
\usepackage{bm}

\newcommand{\pdag}{{\phantom{\dagger}}}

\begin{document}


\title{
Non-equilibrium magnetization dynamics of ferromagnetically coupled Kondo spins
}

\author{Andreas Hackl}
\author{Matthias Vojta}
\affiliation{Institut f\"ur Theoretische Physik, Universit\"at zu K\"oln, Z\"ulpicher Str. 77, 50937 K\"oln, Germany}

\author{Stefan Kehrein}
\affiliation{Arnold Sommerfeld Center for Theoretical Physics, Department f\"ur Physik,
Ludwig-Maximilians-Universit\"at M\"unchen}

\date{\today}

\begin{abstract}
An analytical description of non-equilibrium phenomena in interacting quantum systems is
rarely possible. Here we present one example where such a description can be achieved, namely the ferromagnetic Kondo model. In
equilibrium, this model is tractable via perturbative renormalization-group techniques.
We employ a recently developed extension of the flow-equation method to calculate the
non-equilibrium decay of the local magnetization at zero temperature. The flow equations
admit analytical solutions which become exact at short and long times, in the latter case
revealing that the system always retains a memory of its initial state.
\end{abstract}

\maketitle


\section{\label{intro} Introduction}

Stimulated by the experimental developments in the fields of ultracold gases and of
nanosystems like quantum dots, non-equilibrium phenomena in interacting many-particle
systems are a fascinating field of current research. In the theoretical treatment, a
number of conceptual and methodological problems arise. Some are related to the facts that many
interesting phenomena are not perturbatively accessible even in equilibrium, and that
moreover the limit of a perturbation being small does not necessarily commute with the
long-time limit. Consequently, the majority of works on non-equilibrium physics in
interacting systems rely on numerical methods. However, those come with their own
complications, and analytically tractable test cases are desirable.

It is the purpose of this paper to provide one such test case: We shall study the
non-equilibrium spin dynamics in the ferromagnetic Kondo model,
described by the Hamiltonian
\begin{equation}
H = \sum_{\vec{k}\sigma} \varepsilon_{\vec{k}} c_{\vec{k}\sigma}^\dagger c_{\vec{k}\sigma}^\pdag
+ J^{\parallel } S^z s^z(0) + J^{\perp} [S^x s^x(0) + S^y s^y(0)]
\label{h1}
\end{equation}
in standard notation, with $S^\alpha$ and $s^\alpha(0)$ being the components of the impurity
spin and the conduction electron spin density at the impurity site, and
$J^\|$, $J^\perp$ the exchange couplings, with $|J^\perp| \leq -J^\|$.
This model has the advantage that its equilibrium physics is well understood
and can be accessed by perturbative methods, although bare perturbation theory needs to
be re-summed using renormalization group (RG) techniques.
Here we shall be interested in the decay of a spin prepared in a pure state,
say $|\uparrow\rangle$, after the coupling to the bath is switched on at a time $t\!=\!0$.
To make controlled calculations, we shall focus on the case of small Kondo couplings
$|J^\||$, $|J^\perp| \ll D$ where $2D$ is the conduction-electron bandwidth.
While the equilibrium physics of Eq.~\eqref{h1} is that of an asymptotically free spin, the
non-equilibrium problem is far from trivial. In particular, it is not obvious whether the
expectation value $\langle S^z \rangle(t)$ relaxes to zero:
The equilibrium flow of the spin-flip coupling towards zero is slow (i.e. logarithmic),
which suggests that spin-flip events occur even at long times. In case the spin relaxes to a
non-zero value, the follow-up question is whether this value is identical to the
equilibrium expectation value $\langle S^z \rangle_{\rm eq}$ obtained in the presence of
an infinitesimal magnetic field.

In order to tackle this problem, we shall employ a non-equilibrium extension of the
flow-equation method, originally developed by Wegner \cite{Wegner1994}
and independently -- in the field of high energy physics -- by 
G{\l}azek and Wilson. \cite{GlazekWilson}
For equilibrium settings, this method utilizes continuous unitary transformations to
successively eliminate off-diagonal terms in the Hamiltonian, starting from the highest
energies. Although somewhat similar in spirit to RG techniques,
the flow-equation method has the advantage of diagonalizing, instead of eliminating,
the high-energy part of the Hamiltonian, such that the entire system is
described at any stage of the transformation.
The flow-equation method can be efficiently employed to describe the non-equilibrium time
evolution of observables:\cite{Hackl2007,Hackl2008}
The Hamiltonian is unitarily transformed into a diagonal form, together with the
observables of interest. In this representation, the time evolution can be determined exactly.
Finally, the result is transformed back using the inverse of the flow-equation
transformation. In this way, the accumulation of errors in the long-time limit is
prohibited, and secular terms do not occur.
In Refs.~\onlinecite{Hackl2007,Hackl2008}, this method was applied to dissipative
quantum impurity models, namely the dissipative harmonic oscillator and the ohmic
spin-boson model, and excellent agreement with available analytical and numerical results
was obtained.

\subsection{Summary of results}

In the following, we summarize our main results obtained in the limit of weak coupling,
distinguishing the cases of isotropic and anisotropic Kondo couplings.
Most importantly, in both cases the relaxation of
$\langle S^z (t)\rangle$ as $t\to\infty$ is neither to zero nor to the equilibrium value
$\langle S^z \rangle_{\rm eq}$. Instead, the asymptotic value is given by
$1/2 - \langle S^z (t\to\infty) \rangle = 2(1/2-\langle S^z \rangle_{\rm eq})$, i.e., the
spin polarization is reduced twice as much compared to the equilibrium case.

The initial (short-time) decay of $\langle S^z (t)\rangle$ is set by $J^\perp$ (independent of $J^\parallel$),
$\langle S^z(t)\rangle=\frac{1}{2}\bigl[ 1-(2J^\perp Dt)^2\bigr]$.\cite{unitnote}
In the isotropic case, the long-time decay towards the asymptotic value is logarithmic,
$\langle S^z (t)\rangle - \langle S^z (t\to\infty)\rangle \propto 1/\ln t$.
In contrast, for anisotropic couplings with $J^\| < -|J^\perp|$ the logarithms in the
long-time limit are replaced by power laws,
$\langle S^z (t)\rangle - \langle S^z (t \to \infty)\rangle \propto
t^{2\tilde{j}^\parallel}$, where $\tilde{j}^\parallel<0$ is the dimensionless
fixed-point value of $J^{\| }$.
These results do not come unexpected, considerung the equilibrium RG flow for the
ferromagnetic Kondo problem, which is characterized by logarithmic (power-law) flow in
the isotropic (anisotropic) case.

Below, we shall give explicit expressions for $\langle S^z (t)\rangle$ for all times,
Eqs.~\eqref{resultiso} and~\eqref{resultaniso},
and discuss in detail their evaluation in both the short-time and long-time limits.

\subsection{Relation to earlier work}

Our work adresses both recent methodological developments
and fundamental theoretical questions. Applications of the flow-equation
approach to interaction quenches in quantum many-body systems have been
applied in several recent works studying dissipative quantum systems
\cite{Hackl2007,Hackl2008} and interaction quenches in the Hubbard
model.\cite{Moeckel2008,MoeckelKehrein_AnnPhys}
These works relied on either numerical implementations or
analytical approximations restricted to finite time scales.
Previous numerical results exist for the real-time 
dynamics of the {\em antiferromagnetic} Kondo model,
which has also been treated with the flow equation 
method \cite{LobaskinKehrein}.
The ferromagnetic Kondo model has been discussed numerically
with the time-dependent Numerical Renormalization Group (TD-NRG) 
method \cite{Anders_KM} at finite temperatures \cite{Anders_SB} or in the context of the
underscreened Kondo effect.\cite{Roosen08} These works did not identify
the asymptotic long-time behaviour of this problem at $T=0$.
This behaviour has been adressed by a comparison
of the flow equation approach and the TD-NRG in a recent letter.\cite{hackl_prl}
The asymptotic long-time tails have been clarified by the analytical asymptotics
of the flow-equation approach, to be described in more detail in the present paper,
and the numerical implementation of the TD-NRG.

Relaxation of an impurity spin in a host model was already discussed by Langreth and Wilkins,\cite{Langreth}
who developed a theory of spin resonance in dilute magnetic alloys.
Based on the Kadanoff-Baym approach, these authors derived Bloch-like equations
for the paramagnetic resonance, which predict a relaxation
of the disturbed magnetizations to the instantaneous local
equilibrium magnetization. Our work differs in predicting a relaxation
towards a local non-equilibrium magnetization under slightly
different conditions than considered by Langreth and Wilkins.
A perturbative expansion of Kadanoff-Baym equations as used by Langreth
and Wilkins would fail to reproduce our result, since bare perturbation
theory cannot reproduce our asymptotic long-time relaxation laws.

\subsection{Plan of the paper}

The remainder of the paper is organized as follows:
In Sec.~\ref{sec:toy} we start our considerations by studying an exactly solvable toy model
of a spin coupled to two fermions. This model illustrates a number of remarkable features
which will re-appear in the treatment of the Kondo model.
Sec.~\ref{flowequation} describes the flow-equation transformation for the ferromagnetic
Kondo model, including the transformation of the relevant observables.
The explicit calculation of the time-dependent impurity magnetization is subject of
Sec.~\ref{realtime}. The central results, obtained both analytically for short and long times
and numerically from a full solution of the flow equations, are presented in
Sec.~\ref{analytical}. A brief discussion of limitations and applications closes the
paper. Technical details will be relegated to the appendices.

A brief account of our results has been published in Ref.~\onlinecite{hackl_prl},
together with a comparison to numerical data obtained by the TD-NRG method.


\section{Toy model}
\label{sec:toy}

Several interesting features of the non-equilibrium processes governed by  
the ferromagnetic Kondo model can be captured
by a simple exactly solvable toy model consisting of two fermion levels coupled
to a spin $1/2$ impurity spin $\vec{S}$ through an $SU(2)$-symmetric exchange coupling:
\begin{equation}
H=\sum_{\alpha}\left(c_{\alpha}^\dagger c_{\alpha}^\pdag -d_{\alpha}^\dagger d_{\alpha}^\pdag \right)
 + \frac{g}{2}\vec{S} \cdot \sum_{\alpha,\beta}(c_{\alpha}^\dagger+d_{\alpha}^\dagger)\vec{\sigma}_{\alpha \beta}^\pdag (c_\beta^\pdag + 
d_\beta^\pdag ) \ .
\label{toy}
\end{equation}
Due to the finite Hilbert space the real time evolution problem for
an arbitrary initial state becomes exactly solvable. Specifically we look
at $|\psi(t=0)\rangle=|\psi_0\rangle$, where
\begin{equation}
| \psi_0 \rangle \stackrel{\rm def}{=} | 0 \rangle \otimes | \uparrow \downarrow \rangle \otimes | \uparrow\rangle
\end{equation}
denotes an unoccupied c-electron level state $| 0 \rangle$, a doubly occupied d-electron level
state $| \uparrow \downarrow \rangle$ and the spin-up state $| \uparrow\rangle$ of the impurity spin.
Notice that $|\psi_0\rangle$ is the ground state of
$\sum_{\alpha}\left(c_{\alpha}^\dagger c_{\alpha}^\pdag -d_{\alpha}^\dagger d_{\alpha}^\pdag \right)-B\,S^z$,
where $B>0$ shall cause an infinitesimal Zeeman splitting. Later we will generalize the initial state
to be a product state of a Fermi sea with the impurity spin-up state for the dynamics generated by
the actual ferromagnetic Kondo model.

In our toy model we now investigate how the spin expectation value deviates from its initial value
at time $t=0$ due to the dynamics generated by (\ref{toy}), that is we study the observable
\begin{equation}
\hat{O}=S^z -\frac{1}{2} \ .
\end{equation}
Let us denote the exact eigenstates of $H$ with eigenenergies $\tilde E_n$ by $|\tilde\psi_n\rangle$.
Their explicit construction is shown in appendix~\ref{toyalgebra}, in fact one just needs to diagonalize a $3\times 3$-matrix.

\begin{figure}
\includegraphics[clip=true,width=9.0cm]{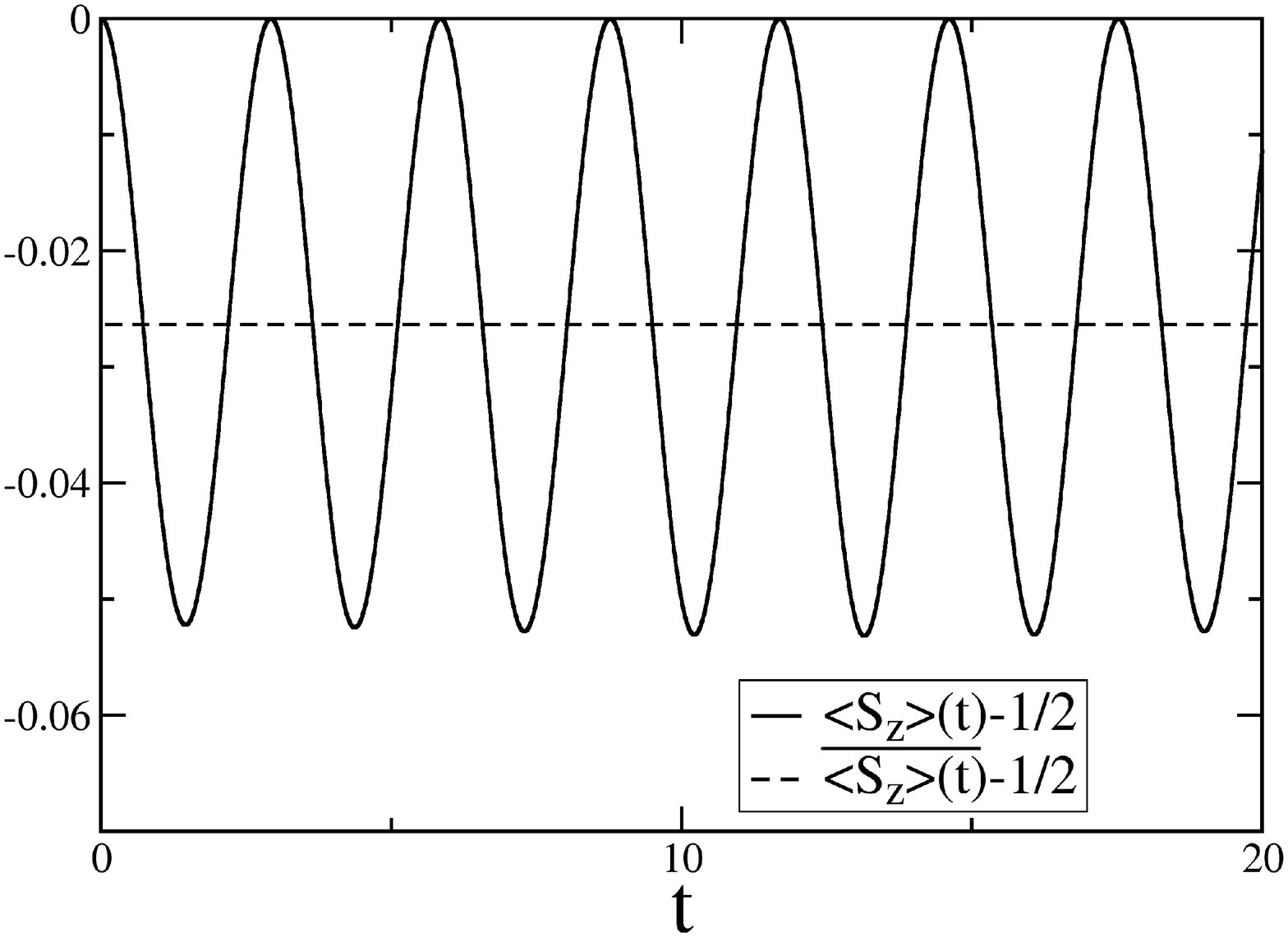}
\caption{\label{toytime}
The time dependence of the expectation value $\langle \psi_0| S^z(t)-\frac{1}{2}\,|\psi_0\rangle$
is shown for $g=-1/2$.
}
\end{figure}

\begin{figure}
\includegraphics[clip=true,width=8.0cm]{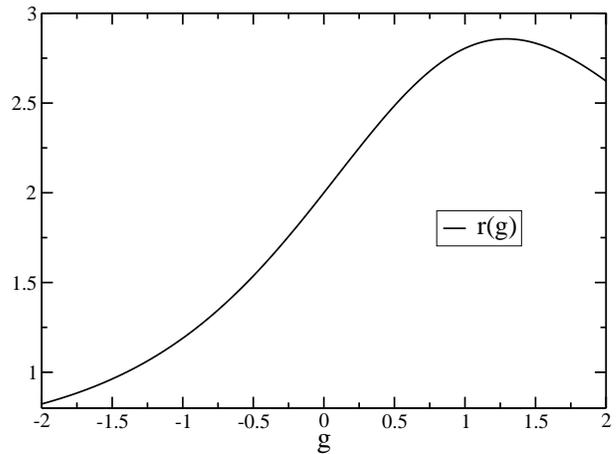}
\caption{\label{toyratio}
We depict the ratio $r(g)=\frac{\overline{\langle\psi_0| S^z(t)-1/2 |\psi_0 \rangle}}{\langle S^z-1/2 \rangle_{eq}}$
as a function  of the coupling strength~$g$. In the weak-coupling limit one finds the universal
ratio $r(0)=2$, see text.
}
\end{figure}

Since the initial state $|\psi_0\rangle$ is not an exact eigenstate, the expectation value of $\hat{O}$
acquires a time dependence given by
\begin{eqnarray*}
O(t)&=&\langle \psi_0|e^{iHt}\,\hat{O}\,e^{-iHt}|\psi_0\rangle \nonumber\\
&=& \sum_{n,n'} u_n^* u_{n'}\, e^{-i(\tilde E_{n'}-\tilde E_n)t}\, \langle \tilde\psi_n|\hat{O}|\tilde\psi_{n'}\rangle \ ,
\end{eqnarray*}
where we have used the decomposition $|\psi_0\rangle = \sum_n u_n |\tilde\psi_n\rangle$.
For an examplary case $O(t)$ is depicted in Fig.~\ref{toytime}: one can clearly observe
oscillations with a finite number of Bohr frequencies $\tilde E_{n'}-\tilde E_n$.
Since there are no accidental degeneracies, the time average is given by
\begin{eqnarray}
\overline{O(t)}&=&
 \lim_{T\rightarrow \infty}\frac{\int_0^T dt \langle \psi_0 | e^{iHt}\,\hat{O}\, e^{-iHt} | \psi_0 \rangle}{T}\nonumber\\
&=& \sum_{n} |u_n|^2 \langle \tilde \psi_n \mid \hat{O} \mid \tilde \psi_n \rangle \ .
\label{mean}
\end{eqnarray}
We compare this average with the ground state expectation value in the interacting system (\ref{toy}):
\begin{equation}
O_{eq}=\langle \tilde\psi_{eq}|\hat{O}|\tilde\psi_{eq} \rangle \ ,
\end{equation}
where $|\tilde\psi_{eq}\rangle$ is the exact eigenstate with the smallest eigenenergy~$\tilde E_n$.
The ratio $\overline{O(t)}/O_{eq}$ is depicted in Fig.~\ref{toyratio} as a function of the coupling strength.
We observe that generically the ratio differs from one, which shows that the time-evolved initial
state $|\psi_0(t)\rangle$ always retains a memory of its initial preparation. While this is not at all
surprising for a finite Hilbert space like in our toy model, we will later see that this observation
is generalized to the ferromagnetic Kondo model in the thermodynamic limit.

The second observation in Fig.~\ref{toyratio} is the universal value~2 in the weak-coupling limit.
The universality of this result is guaranteed by a theorem proven in Ref.~\onlinecite{MoeckelKehrein_AnnPhys}:
Take a discrete quantum system governed by $H=H_0+g\,H_{int}$, where $g\,H_{int}$ is a weak perturbation
for which non-degenerate perturbation theory is possible. Let $|0\rangle$ be the ground state of $H_0$
and $|\tilde 0\rangle$ be the ground state of~$H$. Let $\hat{O}$ be an observable that commutes with $H_0$
and which annihilates $|0\rangle$. Then
\begin{equation}
r(g)\stackrel{\rm def}{=}\frac{\overline{\langle 0|e^{iHt}\,\hat{O}\,e^{-iHt}|0\rangle}}{
\langle \tilde 0|\hat{O}|\tilde 0\rangle}=2+o(g) \ .
\label{defrg}
\end{equation}
One can easily verify that the conditions of this theorem are met in our toy model if we take
\begin{eqnarray*}
H_0&=&\sum_{\alpha} \left(c_{\alpha}^\dagger c_{\alpha}^\pdag -d_{\alpha}^\dagger
d_{\alpha}^\pdag \right) - B\,S^z \\
H_{int}&=&\frac{1}{2}\vec{S} \cdot \sum_{\alpha,\beta}(c_{\alpha}^\dagger+d_{\alpha}^\dagger)\vec{\sigma}_{\alpha \beta}(c_\beta + d_\beta )
\nonumber\\
\end{eqnarray*}
with infinitesimal $B>0$. In appendix~\ref{toyalgebra} we show explicitly how this theorem comes about in our
simple toy model.

Again we will find that this universal factor~2 between the time-averaged expectation
value in the time-evolved initial state and the equilibrium ground state also holds in the
weak-coupling limit of the ferromagnetic Kondo model.


\section{Kondo model and flow-equation transformation}
\label{flowequation}

In this section, we summarize the equilibrium properties of the ferromagnetic Kondo model
which are relevant for the subsequent discussion. We then explain in some detail the
flow-equation treatment of the model, which shall be used in Secs.~\ref{realtime} and
\ref{analytical} to calculate the non-equilibrium magnetization.

\subsection{Definition of the model}

The Kondo Hamiltonian, Eq. \eqref{h1}, can be re-written as
\begin{eqnarray}
H &=&\sum_{\vec{k}\sigma} \varepsilon_{\vec{k}} c_{\vec{k}\sigma}^\dagger c_{\vec{k}\sigma}^\pdag \nonumber\\
&+& \sum_{\vec{k},\vec{k^\prime}} J_{\vec{k^\prime} \vec{k}}^{\| } S^z s_{\vec{k^\prime} \vec{k}}^z \nonumber\\
&+& \sum_{\vec{k},\vec{k^\prime}} J_{\vec{k^\prime} \vec{k}}^{\perp} (S^+ s_{\vec{k^\prime} \vec{k}}^- + S^- s_{\vec{k^\prime}
\vec{k}}^+).
\label{hkondo}
\end{eqnarray}
Here, $S^\pm=S^x \pm i S^y$, and the conduction electron spin density is given by
\begin{equation}
s_{\vec{k^\prime} \vec{k}}^{z,\pm } = \frac{1}{2} \sum_{\alpha,\beta} c_{\vec{k^\prime} \alpha}^\dagger \sigma_{\alpha \beta}^{z,\pm } c_{\vec{k}\beta}^\pdag,
\end{equation}
where $\sigma^\pm=\frac{1}{2}(\sigma^x \pm i\sigma^y)$ derive from the Pauli matrices.
The anisotropic exchange couplings $J_{\vec{k}\vec{k^\prime}}^\perp$ and
$J_{\vec{k}\vec{k^\prime}}^\parallel$ usually describe scattering on an isotropic Fermi
surface, and their momentum dependence can then be safely neglected. In physical results,
they enter proportional to the density of states at the Fermi surface ($\rho_F$), thus we define the
dimensionless couplings $j^\parallel \stackrel{\rm def}{=} \rho_F J^\parallel$
and $j^\perp \stackrel{\rm def}{=} \rho_F J^\perp$.

\subsection{Low-energy fixed points}

The important difference between the ferromagnetic and the antiferromagnetic regime of
the Kondo model can be understood from Anderson's poor man's scaling
analysis.\cite{poormans} In this approach, the band width of the conduction electrons is
progressively reduced by a flowing cutoff energy $\Lambda$ and the renormalized interactions due to the elimination of virtual
excitations to the band edges are calculated perturbatively. As shown by Anderson, this
procedure leads to the scaling equations
\begin{eqnarray}
\frac{dJ^\parallel}{d\ln \Lambda}&=& -\rho_F J^{\perp2} \nonumber\\
\frac{dJ^\perp}{d\ln \Lambda}    &=& -\rho_F J^{\perp}J^\parallel ,
\label{poormans}
\end{eqnarray}
where the scaling trajectory for ferromagnetic
couplings with $|J^\perp|<-J^\parallel$ is given by the conserved quantity 
$J^{\parallel 2}-J^{\perp2}= \text{const}$.  

For antiferromagnetic couplings, more precisely for $J^\| >0 $ or $|J^\perp| > -J^\|$,
the RG flow is driven to strong coupling, i.e., $J^\|\to\infty$ and $|J^\perp|\to\infty$,
and the scaling equations have no low-energy fixed point at finite coupling.
On the level of perturbation theory, logarithmic divergencies $\ln(k_B T/D)$ occur at low temperatures.

In contrast, in the ferromagnetic Kondo model with $|J^\perp| \leq -J^\|$, the perturbative
renormalization of the coupling constant remains controlled in the limit $\Lambda \ll T$
since the coupling monotonously renormalizes to a finite value $J(T)$ in the limit
$\Lambda \rightarrow 0$. At $T=0$ and for anisotropic couplings, the longitudinal
coupling remains finite and the transverse coupling renormalizes to zero
according to the power-law $J^\perp(\Lambda) \propto \Lambda^{\rho_F\sqrt{J_\parallel^2-J_\perp^2}}$. In the
isotropic case, both couplings logarithmically renormalize to zero following
$J(\Lambda)=J/(1+\rho_F J\ln(\Lambda/D))$, and the impurity becomes asymptotically
free at low energies. This important property makes it possible to use the flow equation renormalization
scheme in a perturbatively controlled manner in the following.

\subsection{Flow-equation method}

Let us briefly review the basic ideas of the flow-equation approach
(for more details see Ref. \onlinecite{Kehrein_STMP}). A many-body Hamiltonian~$H$
is diagonalized through a sequence of infinitesimal unitary transformations with
an anti-hermitean generator $\eta(B)$,
\begin{equation}
\frac{dH(B)}{dB}=[\eta(B),H(B)] \ ,
\label{eqdHdB}
\end{equation}
with $H(B\!=\!0)$ the initial Hamiltonian.
The ``canonical" generator \cite{Wegner1994} is the commutator of the
diagonal part~$H_{0}$ with the interaction part~$H_{\rm int}$ of the Hamiltonian,
$\eta(B)\stackrel{\rm def}{=}[H_{0}(B),H_{\rm int}(B)]$. Under rather
general conditions the choice of the canonical generator leads to an increasingly
energy-diagonal Hamiltonian $H(B)$, where interaction matrix elements with
energy transfer~$\Delta E$ decay like $\exp(-B\,\Delta E^{2})$. For $B\rightarrow\infty$
the Hamiltonian will be energy-diagonal and we denote parameters and operators in
this basis by~$\tilde{~}$, e.g. $\tilde H=H(B\!=\!\infty)$.

The key problem of the flow equation approach is generically the generation of higher order interaction
terms in Eq. (\ref{eqdHdB}), which makes it necessary to truncate the scheme in some order of a suitable systematic
expansion parameter (usually the running coupling constant).
Still, the differential nature of the approach
makes it possible to deal with a continuum of energy
scales and to describe non-perturbative effects. This has
led to numerous applications of the flow-equation method
where one utilizes the fact that the Hilbert space is not
truncated as opposed to conventional scaling methods.

\begin{figure}
\includegraphics[clip=true,width=3.2cm,angle=90]{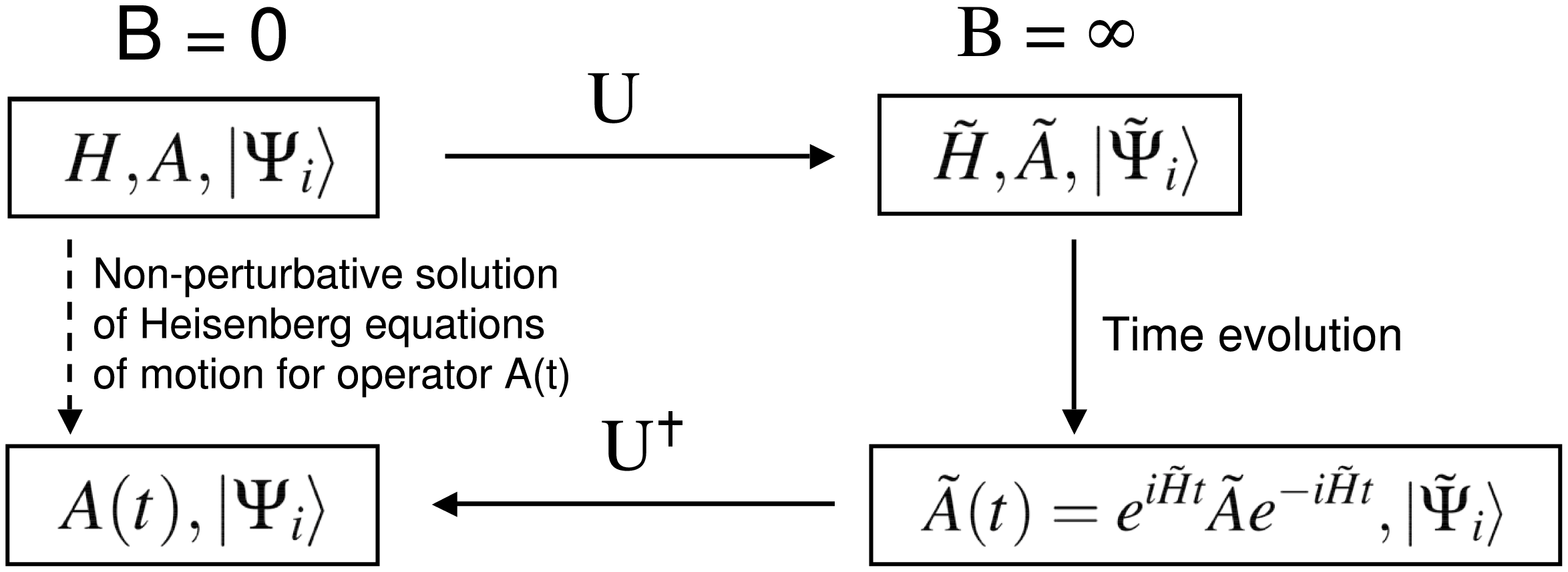}
\caption{\label{figrealtimefeq}
The forward-backward transformation scheme induces
a non-perturbative solution of the Heisenberg equations of motion for an operator.
$U$ denotes the full unitary transformation
that relates the $B=0$ to the $B=\infty$ basis.\cite{footnotefullU}}
\end{figure}

In Refs. \onlinecite{Hackl2007,Hackl2008}, these features have been utilized to treat the real-time
evolution of observables under non-equilibrium conditions. The general
setup is described by the diagram in Fig.~\ref{figrealtimefeq}, where $|\psi_i \rangle$
is some non-thermal initial state whose time evolution
one is interested in. However, instead of following its full
time evolution it is more convenient to study the real
time evolution of a given observable $A$ that one is
interested in. This is done by
transforming the observable into the diagonal basis in
Fig.~\ref{figrealtimefeq} (forward transformation):
\begin{equation}
\frac{d\hat{O}}{dB}=[\eta(B),\hat{O}(B)]
\label{observable}
\end{equation}
with the initial condition $\hat{O}(B\!=\!0) = A$. The central
observation is that one can now solve the real time
evolution with respect to the energy-diagonal $\tilde H$ exactly,
thereby avoiding any errors that grow proportional to
time (i.e., secular terms): this yields $\tilde A(t)$. Now since
the initial quantum state is given in the $B\!=\!0$ basis,
one undoes the basis change by integrating Eq. (\ref{observable}) from
$B\rightarrow\infty$ to $B\!=\!0$ (backward transformation) with the
initial condition $\hat{O}(B\rightarrow\infty) = \tilde A(t)$. One therefore
effectively generates a new non-perturbative scheme for solving
the Heisenberg equations of motion for an operator,
$A(t) = e^{iHt} A(0) e^{-iHt}$.

\subsection{Flow equations for the Hamiltonian}

It is a straightforward calculation to apply the flow equation technique in the outlined
manner to the ferromagnetic Kondo Hamiltonian.\cite{Kehrein_STMP,KehreinKMV}
For the sake of simplicity, we consider the case of $S=1/2$ here and give a generalization
to arbitrary spin $S$ in appendix~\ref{highS}. To start with, we split up the flowing Kondo Hamiltonian as
\begin{equation}
H(B)=H_0 + H_{\text{int}}(B)
\end{equation}
where $H_0=\sum_{\vec{k} \sigma} \varepsilon_{\vec{k}} c_{\vec{k}\sigma}^\dagger c_{\vec{k}\sigma}^\pdag$,
and $H_{\text{int}}(B)$ describes the interaction with the spin-1/2 degree of freedom $\vec{S}$,
\begin{eqnarray}
H_{\text{int}} (B) &=& \sum_{\vec{k^\prime} \vec{k}} J_{\vec{k^\prime}\vec{k}}^{\| }(B) S^z :s_{\vec{k^\prime} \vec{k}}^z: \nonumber\\
&+& \sum_{\vec{k^\prime}\vec{k}} J_{\vec{k^\prime} \vec{k}}^{\perp}(B) (S^+ :s_{\vec{k^\prime}\vec{k}}^-: + S^- :s_{\vec{k^\prime} \vec{k}}^+: ) \ .
\label{ansatzh}
\end{eqnarray}
Normal ordering of the fermion operators has been denoted by $: \quad :$. It is used in order to safely truncate
normal-ordered terms with respect to a given reference state from the ansatz Eq.~\eqref{ansatzh}. These
truncated terms are in addition proportional to higher powers of the coupling $J$ compared to the
leading order of the calculation.\cite{Kehrein_STMP} In the following,
we will use normal ordering w.r.t. the equilibrium density matrix of the non-interacting Fermi sea,
but other choices could be used as well.\cite{Kehrein_STMP,Hackl2007}
According to the ansatz Eq.~\eqref{ansatzh}, we will calculate the flow of the Hamiltonian only
to first order in the coupling $J$. It would be straightforward to formulate
an ansatz that also contains coupling terms proportional to higher orders in $J$, which
are necessarily generated during the flow. For our purpose, it will be sufficient to
work only to lowest nontrivial order, which includes only interaction terms of
first order in the coupling $J$.
The canonical generator is immediately obtained from  $[H_0, H_{\text{int}}(B)]$,
\begin{eqnarray}
\eta (B)&=&[H_0,H_{\text{int}}(B)] \nonumber\\
&=&\sum_{\vec{k^\prime}\vec{k}} \biggl( \eta_{\vec{k^\prime} \vec{k}}^\parallel (B):S^z s_{\vec{k^\prime}\vec{k}}^z: \nonumber\\
&+& \eta_{\vec{k^\prime}\vec{k}}^\perp (S^+ :s_{\vec{k^\prime}\vec{k}}^-: + S^- :s_{\vec{k^\prime}\vec{k}}^+: ) \biggr),
\label{generator}
\end{eqnarray}
with
\begin{eqnarray}
\eta_{\vec{k^\prime}\vec{k}}^{\parallel}(B)&\stackrel{\rm def}{=}&(\varepsilon_{\vec{k^\prime}}-\varepsilon_{\vec{k}} ) J_{\vec{k^\prime} \vec{k}}^{\parallel}(B) \nonumber\\
\eta_{\vec{k^\prime}\vec{k}}^{\perp}(B)&\stackrel{\rm def}{=}&(\varepsilon_{\vec{k^\prime}}-\varepsilon_{\vec{k}} ) J_{\vec{k^\prime} \vec{k}}^{\perp}(B) \ .
\end{eqnarray}
By comparing coefficients on both hand sides of the differential equation $\frac{dH}{dB}=[\eta(B),H(B)]$,
one immediately finds the flow equations for the coupling constants:\cite{Kehrein_STMP}
\begin{eqnarray}
\frac{dJ_{\vec{k^\prime} \vec{k}}^\parallel}{dB}&=&-(\varepsilon_{\vec{k^\prime}}-\varepsilon_{\vec{k}} )^2 J_{\vec{k^\prime} \vec{k}}^\|   \nonumber\\
&+& \sum_{\vec{q}} (2 \varepsilon_{\vec{q}} - \varepsilon_{\vec{k^\prime}} -\varepsilon_{\vec{k}} ) J_{\vec{k^\prime} \vec{q}}^\perp J_{\vec{q}\vec{k}}^\perp \nonumber\\
&\times & (\frac{1}{2}-n(\vec{q})) + \mathcal{O}(J^3)\nonumber\\
\frac{dJ_{\vec{k^\prime} \vec{k}}^\perp}{dB}&=&-(\varepsilon_{\vec{k^\prime}}-\varepsilon_{\vec{k}} )^2 J_{\vec{k^\prime} \vec{k}}^\perp   \nonumber\\
&+& \sum_{\vec{q}} (2 \varepsilon_{\vec{q}} - \varepsilon_{\vec{k^\prime}} -\varepsilon_{\vec{k}} )\nonumber\\
&\times & \frac{1}{2}\bigl( J_{\vec{k^\prime} \vec{q}}^\parallel J_{\vec{q}\vec{k}}^\perp + J_{\vec{k^\prime} \vec{q}}^\perp J_{\vec{q}\vec{k}}^\parallel  \bigr) \bigl(\frac{1}{2}-n(\vec{q})\bigr) \nonumber\\
&+& \mathcal{O}(J^3)
\label{feqH} \ .
\end{eqnarray}
The initial conditions for the exchange interactions are $J_{\vec{k^\prime} \vec{k}}^\parallel(B=0) \equiv J^\parallel/N$
and $J_{\vec{k^\prime} \vec{k}}^\perp(B=0) \equiv J^\perp/N$, where $N$ is the number of band states.
In the flow equations (\ref{feqH}), the Fermi distribution function
$n(\vec{k})=1/(1+\exp(\varepsilon_{\vec{k}} / T ))$ has been introduced, which enters the
flow equations as a contraction generated by our normal-ordering procedure.
Since the couplings  $J_{\vec{k^\prime} \vec{k}}^\perp$ flow to zero in the limit $B\rightarrow \infty$,
the fixed point of the flow-equation renormalization of $H$ is
\begin{eqnarray}
H(B=\infty)&=&\sum_{{\bf k}\sigma}\epsilon_{{\bf k}\sigma} c_{{\bf k}\sigma}^\dagger c_{{\bf k}\sigma}^\pdag \nonumber\\
&+&\sum_{{\bf k}\sigma} J_{{\bf k}{\bf k}}^\parallel(B=\infty) 
S^z \sigma c_{{\bf k}\sigma}^\dagger c_{{\bf k}\sigma}^\pdag \ .
\end{eqnarray}
For a discussion of the magnetization curve $\langle S^z(t) \rangle$, we can neglect the 
term $\sum_{{\bf k}\sigma} J_{{\bf k}{\bf k}}^\parallel(B=\infty)S^z \sigma c_{{\bf k}\sigma}^\dagger c_{{\bf k}\sigma}^\pdag$, since
the renormalized couplings $J_{{\bf k}{\bf k}}^\parallel(B=\infty)$ shift the fermionic energy levels
proportional to the inverse number of band states ($1/N$) and will drop out of the 
impurity magnetization curve $\langle S^z(t)\rangle$ in the thermodynamic limit 
$N\rightarrow \infty$. It is therefore possible to exploit the noninteracting form 
$H(B=\infty)=\sum_{{\bf k}\sigma}\epsilon_{{\bf k}\sigma} c_{{\bf k}\sigma}^\dagger c_{{\bf k}\sigma}^\pdag$
by transforming the impurity spin operator $\vec S$ into the same basis representation.

\subsection{Flow equations for the spin operator}

In order to transform the impurity spin operator $\vec S$ into the basis of the diagonal Hamiltonian $H(B=\infty)$,
we need to solve the differential equation
\begin{equation}
\frac{dS^a(B)}{dB}=[\eta(B),S^a (B)],
\end{equation}
with $a=(x,y,z)$ and the initial condition $S^a(B=0)=S^a$.
By considering the commutator $[\eta(B),S^a]$, it is straightforward to work out the following ansatz for the
flowing operator $S^a(B)$
\begin{equation}
S^a(B)=h(B) S^a+ \sum_{\vec{k^\prime}, \vec{k}} \gamma_{\vec{k^\prime} \vec{k} }(B):(\vec{S} \times \vec{s}_{\vec{k^\prime} \vec{k}})^a:,
\label{szb}
\end{equation}
which is accurate to neglected normal-ordered contributions of $\mathcal{O}(J^2)$.
In the following, we transform only the operator $S^z$ which describes the impurity magnetization.
By extending the coefficients in the ansatz of Eq.~\eqref{szb}
by a dependence on time, the real time evolution of the operator $S^z(B)$
can be parameterized. We will make use of this parameterization later on,
and therefore introduce here the following abbreviations for the couplings,
which are related to specific values of the parameters $B$ and $t$:
\begin{displaymath}
h(B,t) = \left\{ \begin{array}{ll}
h(t) & B=0\\
h(B) & t = 0 \\
\tilde{h}(t) & B \rightarrow \infty\\
\tilde{h} & B \rightarrow \infty , t=0 \ . \\
  \end{array} \right.
\end{displaymath}
The differential flow of the ansatz Eq. (\ref{szb}) is readily obtained
from the commutator $[\eta(B),S^z(B)]$ as \cite{Kehrein_STMP}
\begin{widetext}
\begin{eqnarray}
\frac{dh}{dB} &=& \sum_{\vec{k^\prime} \vec{k}} (\varepsilon_{\vec{k^\prime}} -\varepsilon_{\vec{k}}) J_{\vec{k^\prime} \vec{k}}^\perp(B) \gamma_{\vec{k} \vec{k^\prime}}(B)
n(\vec{k^\prime})(1-n(\vec{k}))\nonumber\\
\frac{d\gamma_{\vec{k^\prime} \vec{k}}}{dB} &=& h(B)(\varepsilon_{\vec{k^\prime}} - \varepsilon_{\vec{k}}) J_{\vec{k^\prime} \vec{k}}^\perp (B)
- \frac{1}{4} \sum_{\vec{u}} \biggl( (\varepsilon_{\vec{k^\prime}} - \varepsilon_{\vec{u}} ) J_{\vec{k^\prime} \vec{u}}^\parallel (B)
\gamma_{\vec{u}\vec{k}}(B) + (\varepsilon_{\vec{k}} -\varepsilon_{\vec{u}}) J_{\vec{u}\vec{k}}^\parallel (B) \gamma_{\vec{k^\prime} \vec{u}}(B) \biggr) ( 1- 2n(\vec{u})) \ .
\label{feqsz}
\end{eqnarray}
\end{widetext}
In Sec.~\ref{analytical}, these differential equations will be simplified in order to derive an approximate solution to them.
For a check of approximations, it will be useful that the flowing couplings $h(B)$ and $\gamma_{\vec{k}\vec{k}^\prime}(B)$
are related by the sum rule \cite{Kehrein_STMP}
\begin{eqnarray}
\langle S^z(B)^2 \rangle &=& \frac{1}{4} \bigl( h^2(B) - \sum_{\vec{k} \vec{k}^\prime} \gamma_{\vec{k} \vec{k}^\prime} \gamma_{\vec{k^\prime}\vec{k}} n(\vec{k}^\prime)(1-n(\vec{k})) \bigr) \nonumber\\
&=& \frac{1}{4} + \mathcal{O} (J^2) \ ,
\label{sumrule}
\end{eqnarray}
where $\langle S^z(B)^2 \rangle$ is evaluated with respect to the
non-interacting Fermi sea at equilibrium and a polarized impurity spin.
This sum rule can be easily proven by differentiating it with respect to $B$
and inserting the flow equations (\ref{feqsz}).


\section{Time-dependent magnetization}
\label{realtime}

Since a ferromagnetically coupled Kondo spin is asymptotically free at low energies, its
equilibrium properties are only perturbatively renormalized in comparison to a free spin.
In the case of isotropic couplings, it has been shown by Abrikosov and Migdal~\cite{Abrikosov} by
diagrammatic means that for general spin quantum number $S$ the equilibrium magnetization
of the impurity ($M=g_i\mu_B\langle S^z\rangle$) in a magnetic field $B$, caused by the Zeeman term
\begin{equation*}
H_{Zeeman}=-g_i \mu_B S^z B
\end{equation*}
is renormalized to one-loop order as
\begin{equation}
M=g_i\mu_B S\left(1-\frac{2(J\rho)^2 x/2}{1+(J\rho )x}\right) \ .
\label{magnetization}
\end{equation}
Here, we introduced $x=\ln(\varepsilon_F/(g_i\mu_B B))$ with the Fermi energy
$\varepsilon_F$, the gyromagnetic ratio $g_i$ of the impurity and the Bohr magneton
$\mu_B$. Spin-flip scattering effectively reduces the magnetization but is not able to
screen the spin completely. If otherwise an impurity spin is decoupled from the
conduction electrons and thus its magnetization is completely polarized, a subsequent
coupling to the conduction band will lead to spin-flip scattering and the impurity magnetization
will start to decrease as a function of time. In order to analyze this behaviour quantitatively, we assume in
the following that the complex of impurity and conduction electrons is prepared
in the product inital state
\begin{equation}
|\psi\rangle = |\uparrow \rangle \otimes |FS\rangle \ ,
\end{equation}
where $|FS\rangle$ is the non-interacting Fermi sea at equilibrium. 
We assume furthermore that impurity and conduction electrons become coupled
instantaneously at the time $t=0$.

Using a perturbative expansion of the Heisenberg equation of motion for the magnetization
operator $S^z$, it is straightforward to calculate the perturbative time evolution of the
magnetization as (with $\langle S^z \rangle=\frac{1}{2}$) \cite{Anders_SB}
\begin{eqnarray}
\langle S^z(t)\rangle &=& \langle S^z\rangle + \langle S^z\rangle (iJ^\perp )^2 \int_{-\infty}^{\infty} d\varepsilon \int_{-\infty}^{\infty}
d \varepsilon^\prime \rho(\epsilon) \rho(\varepsilon^\prime) \nonumber\\
&\times & f(\varepsilon^\prime ) (1-f(\varepsilon )) \Bigl[ \frac{1-\cos((\varepsilon-\varepsilon^\prime)t)}{(\varepsilon-\varepsilon^\prime)^2}\Bigr] \nonumber\\
&+& \mathcal{O}(J^3),
\label{shorttime}
\end{eqnarray}
with the Fermi function $f(\varepsilon)$ and the conduction electron density of states $\rho(\epsilon)=\sum_{\vec{k}}\delta(\epsilon -\epsilon_{\vec{k}})/N$,
where $N$ is the number of band states.
This result neglects contributions of $\mathcal{O}(J^3)$ in coupling strength. It is a well known
property of the Kondo model that renormalization of the couplings $J^\perp$ by
$J^\parallel$ sets in precisely at third order in $J$.\cite{Hewson} At sufficiently large
time scales, the low-energy couplings are expected to dominate the magnetization dynamics, and the renormalization of
spin-flip scattering will become important. Beyond short times, the perturbative result
of Eq. (\ref{shorttime}) is therefore expected to break down.

\begin{figure}
\includegraphics[clip=true,width=8.0cm]{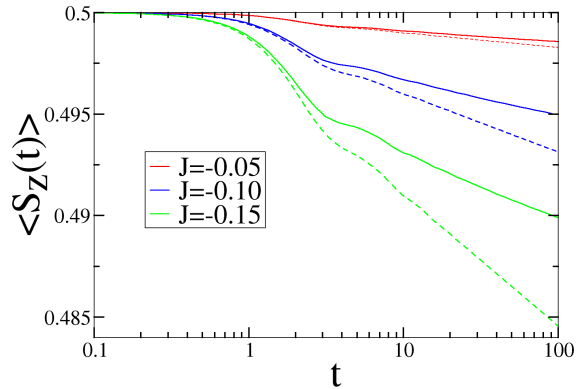}
\caption{\label{isoshort}
We show $\langle S^z(t) \rangle$ for isotropic coupling $j$ and short times.
For very short times $t \ll D^{-1}$, the perturbative result from Eq.~(\ref{shorttimeresult}) (dashed line) is asymptotically coinciding with
a fully numerical solution of the fkow equation~\eqref{feqsz} (full lines). The renormalization of $J^\perp$ by $J^\parallel$
sets in at energy scales $E<D$, leading to a reduction of spin-flip scattering.
Beyond the short-time regime $t \ll D^{-1}$, the magnetization relaxes therefore slower than predicted by the perturbative result.
}
\end{figure}

In the limit $T\rightarrow 0$ and for a flat band with density of states $\rho_F=1/(2D)$,\cite{foot1} the integrals can be rewritten as
\begin{equation}
\langle S^z(t)\rangle = \langle S^z\rangle - \langle S^z\rangle \int_0^{2D} \frac{k(x)}{x^2} j^2 (1-\cos(xt)) dx,
\label{shorttimeint}
\end{equation}
where we introduced the function
\begin{equation}
k(x) \stackrel{\rm}{=} \biggl\{
\begin{array}{l}
x, ~~~~~~~~~x \leq D\\
2D-x, ~x > D.
\end{array}
\label{kdefine}
\end{equation}
This integral can be solved analytically with the result
\begin{equation}
\langle S^z(t)\rangle = \langle S^z\rangle - \langle S^z\rangle (2 J^\perp )^2 [G(2Dt)-2G(Dt)]
\label{shorttimeresult}
\end{equation}
where the function $G(x)$ is defined by the series expansion
\begin{equation}
G(x)= \sum_{l=1}^{\infty} \frac{(-1)^{l+1}}{(2l)!2l(2l-1)}x^{2l}.
\end{equation}
As argued above, Eq.~\eqref{shorttimeresult} fails to predict the correct saturation behaviour of the magnetization,
see also Figs~\ref{isoshort} and~\ref{anisoshort}.

To improve upon this unsatisfactory result, we perform
the perturbative solution of the Heisenberg equation for the operator $S^z$ in
the basis given by the diagonalized form of the Hamiltonian, see Fig.~\ref{figrealtimefeq}.
In several previous applications,\cite{Hackl2008,Hackl2007,Moeckel2008}
this approach has been used to obtain the real-time evolution of physical
observables well beyond the perturbative short-time regime.

In the corresponding new basis representation, the solution to the Heisenberg equation
of motion for $S^z$ is readily
obtained from $\tilde{S}^z (t)=e^{i\tilde{H} t} \tilde{S}^z e^{-i\tilde{H} t}$. Since the
operator $S^z$ is a conserved quantity in this basis, in the ansatz Eq.~\eqref{szb} only
the coefficients $\tilde{\gamma}_{\vec{k}\vec{k}^\prime}$ will obtain time-dependence, with the
result
\begin{eqnarray}
\tilde{S}_z(t) &=& \tilde{h} S^z \nonumber\\
         &+& \sum_{\vec{k^\prime}, \vec{k}} \tilde{\gamma}_{\vec{k^\prime} \vec{k}} (t) :\biggl( (S^+ s_{\vec{k^\prime} \vec{k}}^- + S^- s_{\vec{k}\vec{k^\prime}}^+\biggr): \nonumber\\
\tilde{\gamma}_{\vec{k^\prime} \vec{k}}(t)&=& \tilde{\gamma}_{\vec{k^\prime} \vec{k}} e^{it(\varepsilon_{\vec{k^\prime}}-\varepsilon_{\vec{k}})}.
\label{tildeszt}
\end{eqnarray}
The inversion of the unitary transformation (denoted by $U$, see Fig.
\ref{figrealtimefeq}) will yield the effective non-perturbative solution of the
Heisenberg equation of motion for the operator $S^z$, which we write as
\begin{eqnarray}
S^z(t)&=&h(t) S^z+ \sum_{\vec{k^\prime}, \vec{k}} \gamma_{\vec{k^\prime} \vec{k}} (t) :\biggl( S^+ s_{\vec{k^\prime} \vec{k}}^- + S^- s_{\vec{k}\vec{k^\prime}}^+\biggr): \nonumber\\
&+& \mathcal{O}(J^2).
\label{szt}
\end{eqnarray}
All coefficients in Eq.~(\ref{szt}) are obtained by integrating the flow equations (\ref{feqsz})
from $B \rightarrow \infty$ to $B=0$, with the initial condition in $B \rightarrow \infty$ posed
by the operator $\tilde{S}^z(t)$, Eq. (\ref{tildeszt}). From Eq. (\ref{szt}), the magnetization
$\langle S^z(t)\rangle$ can be readily read off as $\langle S^z(t)\rangle=  h(t)/2$.
It remains to determine the coefficient function $h(t)$, for which an
analytical result can be derived in a limit of long times.


\section{\label{analytical} Analytical results for the magnetization}

It is too ambitious to solve the flow equations (\ref{feqH}) and (\ref{feqsz})
analytically as they stand, but much progress can be made by choosing a suitable ansatz
for the flow of the coupling constants $J^\parallel_{\vec{k} \vec{k^\prime}}$ and
$J^\perp_{\vec{k} \vec{k^\prime}}$. Several suggestions along this line have been given
in Ref. \onlinecite{Kehrein_STMP}, chapter 2. In the limit of exchange couplings $J_{\vec{k}
\vec{k^\prime} }^{\perp , \parallel}$ close to the Fermi surface, meaning
$|\varepsilon_{\vec{k}}|/D  , |\varepsilon_{\vec{k^\prime}}|/D \ll 1$, the flow of these
couplings can be formally parameterized as
\begin{eqnarray}
J_{\vec{k} \vec{k^\prime}}^{\perp , \parallel}(B) \stackrel{\rm def}{=} \frac{J_{IR}^{\perp, \parallel}(B)}{N} e^{-B(\varepsilon_{\vec{k}}-\varepsilon_{\vec{k^\prime}})^2}.\
\label{infrared}
\end{eqnarray}
We call this parameterization of the flow the {\sl infrared parameterization} with the {\sl
infrared couplings} $J_{IR}^{\perp, \parallel}(B)$, since it is asymptotically exact for
$\varepsilon_{\vec{k}}, \varepsilon_{\vec{k^\prime}}=0$. The flow of the couplings
$J_{IR}^{\perp, \parallel}(B)$ is identically to that of $J_{k_F k_F}^{\perp ,
\parallel}(B)$, which are the couplings at the Fermi surface.\cite{Kehrein_STMP} Since during the flow, high
energy couplings with energy transfer $\Delta E$ are eliminated according to the relation
$\Delta E \propto B^{-1/2}$, the flow  of the low-energy couplings can be shown to be
negligible for $B\leq D^{-2}$. \cite{Kehrein_STMP}
By suitably adjusting the initial conditions to $J^\parallel(B=D^{-2})=J^\parallel$
and  $J^\perp(B=D^{-2})=J^\perp$ the flow of the dimensionless
coupling constants $j^{\parallel}=\rho_F J_{IR}^{\perp}$ and 
$j^{\perp}=\rho_FJ_{IR}^{\parallel}$ coincides then exactly with the scaling equations
(\ref{poormans}), where the flowing bandwidth $\Lambda$ is identified as
$\Lambda=B^{-1/2}$.

In addition, it is possible to simplify the flow equations~\eqref{feqsz} for the spin operator $S^z$
by neglecting the contribution to the derivatives $\frac{d\gamma_{\vec{k^\prime} \vec{k}}}{dB}$
of second order in the couplings $J^\parallel$ and $J^\perp$, setting
\begin{equation}
\frac{d\gamma_{\vec{k^\prime} \vec{k}}}{dB} = h(B)(\varepsilon_{\vec{k^\prime}} -
\varepsilon_{\vec{k}}) J_{\vec{k^\prime} \vec{k}}^\perp(B) +\mathcal{O}((\varepsilon_{\vec{k^\prime}} - \varepsilon_{\vec{k}})J^2) \ .
\label{treeflow}
\end{equation}
This approximation has to be carefully justified for anisotropic couplings $|J^\perp|<|J^\parallel|$ and $J^\parallel<0$,
since the couplings $J_{\vec{k}^\prime \vec{k}}^\parallel$ contained in the neglected terms 
flow to a finite value at the Fermi surface (given by $\epsilon_{\vec{k}}=\epsilon_{\vec{k}^\prime}$=0),
while $J_{\vec{k^\prime} \vec{k}}^\perp(B)$ will flow to zero. We justify this approximation in detail in appendix~\ref{tree}.

Our approximations so far aim at the behavior of the couplings $\gamma_{\vec{k}\vec{k}^\prime}$ near the Fermi surface,
which therefore are suitable to obtain the long-time asymptotics of the impurity
magnetization. We will show now that these low-energy couplings depend decisively on the
behavior of the scaling equations~\eqref{poormans}, leading to qualitatively different
results in the anisotropic regime vs. the isotropic regime of the exchange couplings.

\subsection{Isotropic regime}

In the following, we restrict ourselves to the isotropic case $J=J^\perp=J^\parallel$.

\subsubsection{Flow of the dimensionless coupling $j$}

As already pointed out before, we will be mainly interested in the behaviour of the flowing coupling
$J_{\vec{k}^\prime \vec{k}}(B)$ for energies close to the Fermi surface. Exactly {\sl at} the Fermi 
surface, the flow of the dimensionless coupling
$j=J\rho_F$ according to the scaling equation (\ref{poormans}) can be integrated exactly,
yielding
\begin{equation}
j(\Lambda)=\frac{j}{1+j \ln(\frac{\Lambda}{D})}.
\label{oneloopflow}
\end{equation}
We conclude that the flow of the infrared coupling is given by $J_{IR}(B)=J/\bigl(1-(j/2)\ln(B D^2)\bigr)$,
using the correspondence $\Lambda=B^{-2}$.

\subsubsection{Couplings of the spin operator}

For a further simplification of the flow equation~\eqref{treeflow}, it is important that the coupling 
$\tilde{h}$ will be a number close to $1$, with corrections
that vanish continuously as $J\rightarrow 0$.
Using this fact, we set $h(B)\equiv 1$ in Eq. (\ref{treeflow}), leading to corrections that become small
for small $J$, as we explain now.

More precisely, this approximation can be justified by a wellknown approximation
for the impurity magnetization in an external magnetic field.
In presence of an infinitesimal Zeeman splitting $B S^z ~,\,B \rightarrow 0^+$, the coupling $\tilde{h}$ describes
the ground state  expectation value of the operator $S^z$ by the relation
$\tilde{h}/ 2= \langle S^z\rangle$. By multiplying $\langle S^z\rangle=\tilde{h}/2$
with the impurity gyromagnetic ratio $g_i$ and the Bohr magneton $\mu_B$, 
the impurity magnetization is given as $M=g_i \mu_B \tilde{h}/2$.
However, this magnetization can be easily read off from (\ref{magnetization}) by taking the limit $B\rightarrow 0^+$,
which is $M=(1/2) g_i \mu_B S(1+ j/2)$, and thus we obtain (for $S=1/2$)
\begin{equation}
\tilde{h}= 1+\frac{j}{2},
\label{hinfty}
\end{equation}
where possibly corrections of $\mathcal{O}(j^2)$ occur. Since $\tilde{h}$ is renormalized only to $\mathcal{O}(J)$,
equation (\ref{treeflow}) can be simplified by setting $h(B) \equiv 1$ and neglecting corrections of $\mathcal{O}(J^2)$.

We now aim at the flow of the couplings $\gamma_{\vec{k^\prime} \vec{k}}(B)$, which can
be evaluated asymptotically in the limits $|\varepsilon_{\vec{k}}| ,
|\varepsilon_{\vec{k^\prime}}| \ll D$ and $(D-\varepsilon_{\vec{k}}) /D \ll 1,
(D-\varepsilon_{\vec{k}^\prime}) /D \ll 1$. First, we concentrate on couplings near the
Fermi surface $(|\varepsilon_{\vec{k}}| , |\varepsilon_{\vec{k^\prime}}| \ll D)$, for which
the infrared parameterization from Eq. (\ref{infrared}) can be used in Eq.
(\ref{treeflow}), and we obtain
\begin{eqnarray}
\gamma_{\vec{k}\vec{k^\prime}}(B) &=& \int_0^B (\varepsilon_{\vec{k}} - \varepsilon_{\vec{k^\prime}} ) \frac{J_{IR}(B^\prime)}{N}e^{-B^\prime(\varepsilon_{\vec{k}}-\varepsilon_{\vec{k}^\prime})^2}
dB^\prime \nonumber\\
&+&\mathcal{O}(J^2)
\label{gammaint}
\end{eqnarray}
According to Eq. (\ref{oneloopflow}), the infrared couplings $J_{IR}(B)$ depend only
logarithmically on $B$, and in the integral (\ref{gammaint}), we can set $J_{IR}(B)
\equiv J_{IR}(B=\bigl(\varepsilon_{\vec{k^\prime}}-\varepsilon_{\vec{k}} \bigr)^{-2}) + \mathcal{O}(J^2)$, leading to the result
\begin{eqnarray}
\gamma_{\vec{k^\prime} \vec{k}}(B)&=&\frac{J_{IR}(B=\bigl(\varepsilon_{\vec{k^\prime}}-\varepsilon_{\vec{k}} \bigr)^{-2})}{N(\varepsilon_{\vec{k^\prime}}-\varepsilon_{\vec{k}})} (1-e^{-B(\varepsilon_{\vec{k^\prime}}-\varepsilon_{\vec{k}})^2})\nonumber\\
 &+&\mathcal{O}(J^2) \nonumber\\
\tilde{\gamma}_{\vec{k^\prime} \vec{k}}&=&\frac{J_{IR}(B=\bigl(\varepsilon_{\vec{k^\prime}}-\varepsilon_{\vec{k}} \bigr)^{-2})}{N(\varepsilon_{\vec{k^\prime}}-\varepsilon_{\vec{k}})} +\mathcal{O}(J^2) \ .
\label{gammaflow}
\end{eqnarray}
Now, it is possible to determine the coupling $\tilde{h}$ as a solution of Eq.~\eqref{feqsz}.
Employing Eq. (\ref{treeflow}) in Eq. (\ref{feqsz}) and setting $h(B)\equiv 1$, we first obtain the formal result
\begin{eqnarray}
\tilde{h}-1 &=& -\frac{1}{2}\sum_{\vec{k}\vec{k^\prime}}\int_0^{\infty} dB \frac{d}{dB}\gamma_{\vec{k}\vec{k^\prime}}^2(B) n(\vec{k^\prime})(1-n(\vec{k})) \nonumber\\
&=& -\frac{1}{2}\sum_{\vec{k}\vec{k^\prime}}\tilde{\gamma}_{\vec{k}\vec{k^\prime}}^2 n(\vec{k^\prime})(1-n(\vec{k})).
\label{hinftyfeq}
\end{eqnarray}
At low temperatures, only the low-energy couplings from Eq. (\ref{gammaint}) contribute, and at zero temperature and using a flat band
with density of states $\rho_F=\frac{1}{2D}$, we can rewrite the momentum sums as an integral. Employing finally the logarithmic scaling behavior of
the infrared couplings $J_{IR}(B=\bigl(\varepsilon_{\vec{k^\prime}}-\varepsilon_{\vec{k}} \bigr)^{-2})$ from Eq. (\ref{poormans}) in Eq. (\ref{hinftyfeq}), we arrive at the result
\begin{eqnarray}
\tilde{h}-1 &=& - \frac{1}{2}\int_0^{2D} dx \frac{k(x)}{x^2}\frac{j^2}{(1+j\ln(\frac{x}{D}))^2}.
\label{hinftyint}
\end{eqnarray}
This integral can be evaluated exactly, and to leading order in $j$ we obtain the result
$\tilde{h}=1+j/2+\mathcal{O}(j^2)$. To leading order in $j$, this result is confirmed  by Eq.
(\ref{hinfty}), which has been originally derived by Abrikosov and
Migdal.\cite{Abrikosov}

\subsubsection{Time-dependent magnetization}

As discussed in Sec.~\ref{realtime}, the time-dependent
magnetization $\langle S^z (t)\rangle$ follows directly from the coefficient $h(t)$.
It is therefore necessary to solve the flow equations (\ref{feqsz}) for time-dependent initial
conditions, which endows the flowing couplings with an additional time-dependence, and we change our notation according to
$h(B) \rightarrow h(B,t)$ and $\gamma_{\vec{k}\vec{k^\prime}}(B) \rightarrow \gamma_{\vec{k}\vec{k^\prime} }(B,t)$.

Nevertheless, it is trivial to
obtain $\gamma_{\vec{k}\vec{k^\prime}}(B,t)$ to leading order in $J$, since the derivatives in Eq.~(\ref{gammaflow}) do not depend on time and can be integrated
in analogy to Eq. (\ref{gammaflow}), with the result
\begin{eqnarray}
\gamma_{\vec{k^\prime} \vec{k} }(B,t) &=& \tilde{\gamma}_{\vec{k^\prime} \vec{k}}(t) + \int_{\infty}^B (\varepsilon_{\vec{k^\prime}}-\varepsilon_{\vec{k}})J_{\vec{k^\prime} \vec{k} }(B^\prime) dB^\prime \nonumber\\
&+& \mathcal{O}(J^2)\nonumber\\
&=& \gamma_{\vec{k^\prime} \vec{k} }(B) +\tilde{\gamma}_{\vec{k^\prime} \vec{k}} (e^{it(\varepsilon_{\vec{k^\prime}}-\varepsilon_{\vec{k}})}-1) \nonumber\\
&+& \mathcal{O}(J^2) \ .
\label{treelevel}
\end{eqnarray}
This expression can be employed in the flow equation~\eqref{feqsz} for the coupling $h(B,t)$, which 
can be integrated in analogy to Eq.~(\ref{hinftyfeq}). The formal result for the coupling $h(t)$ is finally
\begin{eqnarray}
h(t)&=&\tilde{h} + \sum_{\vec{k}\vec{k^\prime}} \int_\infty^0 \frac{d\gamma_{\vec{k^\prime} \vec{k}}}{dB} \tilde{\gamma}_{\vec{k}\vec{k^\prime}}(e^{it(\varepsilon_{\vec{k}} - \varepsilon_{\vec{k^\prime}})}-1)\nonumber\\
&\times& n(\vec{k^\prime})(1-n(\vec{k})) \nonumber\\
&+&  \sum_{\vec{k}\vec{k^\prime}} \int_{\infty}^0 \frac{d\gamma_{\vec{k^\prime}\vec{k}}}{dB} \gamma_{\vec{k}\vec{k^\prime}}(B)n(\vec{k^\prime})\bigr(1-n(\vec{k})\bigl) \nonumber\\
&=& \tilde{h} + \sum_{\vec{k}\vec{k^\prime}} \tilde{\gamma}_{\vec{k}\vec{k^\prime}}^2(e^{it(\varepsilon_{\vec{k}}-\varepsilon_{\vec{k^\prime}})}-\frac{1}{2}) \nonumber\\
&\times& n(\vec{k^\prime})(1-n(\vec{k})) \ .
\label{formalht}
\end{eqnarray}
In appendix~\ref{alternative}, we derive this result in a different way by employing the sum rule from Eq.~(\ref{sumrule}),
thereby further justifying all approximations.
Aiming at the long-time behaviour at $T=0$, we employ Eq.~(\ref{gammaflow}) and rewrite $h(t)$ as an integral, in analogy to Eq.
(\ref{hinftyint}). This yields directly the magnetization $\langle S^z(t)\rangle = h(t)/2$,
\begin{equation}
\langle S^z(t)\rangle = \frac{1}{2}+\frac{1}{2}\int_0^{2D} \frac{k(x)}{x^2} \frac{j^2}{(1+j \ln(\frac{x}{D}))^2} (\cos(xt)-1) dx \ ,
\label{resultiso}
\end{equation}
with $k(x)$ defined in Eq.~\eqref{kdefine}.

\begin{figure}
\includegraphics[clip=true,width=8.0cm]{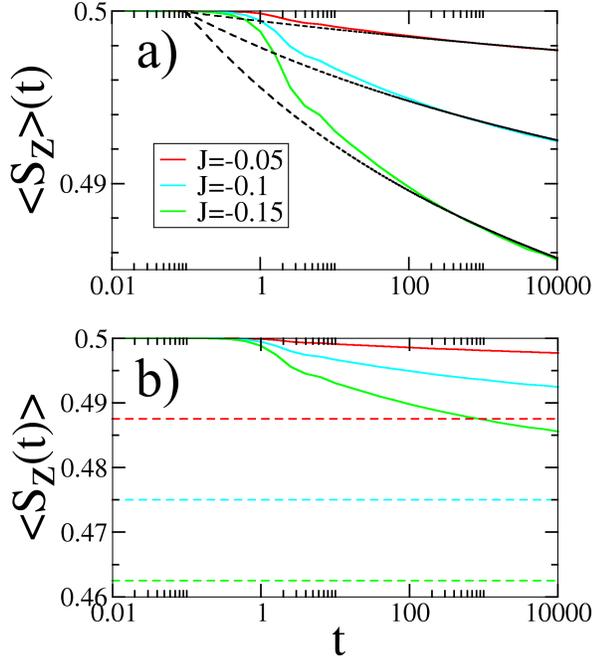}
\caption{\label{isolong} We show $\langle S^z(t) \rangle$ for different isotropic couplings $j$.
In panel a) the full numerical solution of the flow equations (full lines in both panels) is compared to a
fit to the asymptotic analytical behaviour of Eq.~\eqref{isoresult} (dashed lines, fit parameters given in
Table \ref{table1}). Beyond the perturbative short-time regime, described by unrenormalized spin-flip scattering
according to Eq. (\ref{shorttimeresult}), the logarithmic renormalization of the spin-flip scattering
coupling $J^\perp$ sets in. This  leads to a logarithmically slow relaxation of the magnetization,
and the asymptotic value $\langle S^z(t\rightarrow \infty)\rangle=0.5+\frac{j}{2}$ (dashed lines in panel b)) is reached extremely
slowly.
}
\end{figure}

\begin{table}
\caption{\label{table1} For large times, we fitted the full numerical solutions
of the flow equations against the function $0.5\bigl(1+aJ+1/( \ln t +1/(bJ))\bigr)$. For couplings $-j \leq 0.1$,
quantitative agreement with the analytical result $\langle S^z(t)\rangle=0.5\bigl(1/(\ln t-1/j)+ 1/2 +j/2 +\mathcal{O}(j^2)\bigr)$
is very good.}
\begin{ruledtabular}
\begin{tabular}{c|  c c c}
$\boldsymbol{j=j^\perp=j^\parallel}$   & $-0.05$ &  $-0.1$ & $-0.15$\\
\hline
a  & 0.441  & 0.442  & 0.441 \\
b  & 0.418  & 0.400  & 0.382 \\
\end{tabular}
\end{ruledtabular}
\end{table}

For not too low energies, the logarithmic correction $j\ln(x/D)$ can be neglected to leading order in $j$, and the coupling $j$ can be considered
as unrenormalized. In this way, the correct perturbative short-time limit of Eq.~\eqref{shorttimeint} is recovered.
In the long-time limit, $\int_0^{\infty} dx\cos(xt) \dots$ can be replaced by $\int_0^{f(t)/t}dx \dots~$, where $f(t)$ is a function with values of $\mathcal{O}(1)$.
Thus, the long-time tail of the magnetization curve is readily obtained as
\begin{equation}
\langle S^z(t)\rangle = \frac 1 2 \left[ 1/ (\ln t-1/j) + 1 +j +\mathcal{O}(j^2)\right] \ ,
\label{isoresult}
\end{equation}
with the asymptotic value
\begin{equation}
\langle S^z(t\rightarrow \infty)\rangle = 0.5 \bigl(1 +j +\mathcal{O}(j^2)\bigr) \ .
\end{equation}
The steady state magnetization mirrors the behaviour observed in our toy model:
the reduction from full polarization is $j/2$, which is twice the equilibrium value (see Eq.~\ref{hinfty}).

We conclude that Eq.~(\ref{resultiso}) describes both the short-time and the long-time limit
of $\langle S^z(t)\rangle$ exactly. 
This result can be also derived within a more sophisticated ansatz for the flowing couplings
$J^{\perp,\parallel}_{\vec{k}\vec{k^\prime}}(B)$, that parameterizes also the flow of couplings considerably above the Fermi energy, as detailed
in appendix~\ref{diagonal}. 

\subsection{\label{anisosec}Anisotropic regime}

\subsubsection{Flow of the exchange couplings}

Before turning to the dynamical behaviour of the observable
$\langle S^z (t)\rangle$, we recall important scaling properties
of anisotropic exchange couplings $J^\perp$ and $J^\parallel$.
In the ferromagnetic regime, we have $J^\parallel <0$ and a stable fixed point
of the scaling equations \eqref{poormans} exists only if $|J^\parallel| > |J^\perp|$.
In this case, we have
\begin{eqnarray}
\tilde{j}^{\parallel} &=& - \sqrt{j^{\parallel 2} -j^{\perp 2}}\nonumber\\
\tilde{j}^\perp       &=& 0,
\end{eqnarray}
and according to Eq. (\ref{poormans}), the dimensionless transverse coupling decays asymptotically as
$j^\perp(B) = \alpha B^{\tilde{j}^\parallel/2}$, where $\alpha$ is a non-universal
number. Numerical tests show that $\alpha$ is within good accuracy identical to the
coupling $j^\perp$ as long as $|J^\parallel/ J^\perp| \gtrsim 2$. These properties
will be used in the following.

\subsubsection{Flow of the spin operator}

Our discussion of the flowing spin operator starts again with an analysis
of the couplings $\gamma_{\vec{k}\vec{k^\prime}}(B)$ that are generated
during the flow of the observable $S^z$.
As justified in appendix~\ref{tree}, also for anisotropic couplings it is possible to approximate the flow of the
couplings $\gamma_{\vec{k}\vec{k^\prime}}(B)$ by Eq.~\eqref{gammaflow}. In the limit of small energy
differences $\Delta \varepsilon \stackrel{\rm def}{=} |\varepsilon_{\vec{k}}
-\varepsilon_{\vec{k^\prime}}| \ll D$, the integrated flow $\int_0^\infty J_{IR}^\perp(B)dB$ is
dominated by the slow asymptotic decay of $J_{IR}^{\perp}(B)$, and we can set
$J_{\text{IR}}^{\perp}(B)=(\alpha/\rho_F) B^{\tilde{j}^\parallel/2}$,
yielding the approximated couplings
\begin{eqnarray}
\tilde{\gamma}_{\vec{k}\vec{k^\prime}}&=&\int_0^\infty (\varepsilon_{\vec{k}} - \varepsilon_{\vec{k^\prime}}) e^{-B(\varepsilon_{\vec{k}} - \varepsilon_{\vec{k^\prime}})^2}
\frac{J_{IR}^\perp(B)}{N}dB \nonumber\\
&=& \frac{\alpha \Delta \varepsilon}{N\rho_F} \int_0^{\infty} e^{-B \Delta \varepsilon^2} B^{\frac{1}{2}\tilde{j}^\parallel}dB\nonumber\\
&=& \frac{\alpha sgn(\Delta \varepsilon)}{N\rho_F |\Delta \varepsilon|^{1+\tilde{j}^\parallel}} + \mathcal{O}(\alpha \tilde{j}^\parallel) \ .
\label{gammaaniso}
\end{eqnarray}
Since Eq.~(\ref{hinftyfeq}) is valid also for anisotropic exchange couplings, we can again use it
to calculate the renormalized coupling $\tilde{h}$, and by employing the couplings~\eqref{gammaaniso}
in Eq.~\eqref{hinftyfeq}, we obtain the result 
\begin{equation}
\tilde{h}=1-\frac{1}{2}\int_0^{2D} dx  \alpha^2\frac{k(x)}{x^{2+2\tilde{j}^\parallel }} =
1+\frac{\alpha^2}{4\tilde{j}^\parallel}+\mathcal{O}(J^2) \ .
\end{equation}

\subsubsection{Time-dependent magnetization}

\begin{figure}
\includegraphics[clip=true,width=8.5cm]{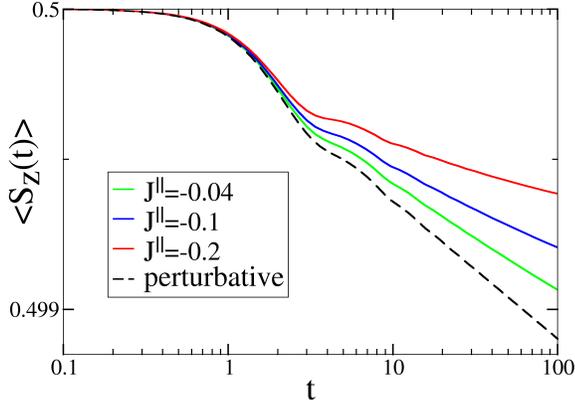}
\caption{\label{anisoshort}
Short-time behaviour of the magnetization for anisotropic couplings from
a fully numerical solution of the flow equations in comparison with the
perturbative result from Eq.~(\ref{shorttimeresult}) (dashed line). The perpendicular coupling is fixed to $J^\perp=-0.04$,
the curve with $J^\parallel=-0.04$ corresponds to the isotropic case.
For increasing $|J^\parallel|$, the spin-flip scattering coupling
$J^\perp$ is stronger renormalized, and the magnetization tends to decay slower than predicted by the unrenormalized
perturbative result.
}
\end{figure}

\begin{figure}
\includegraphics[clip=true,width=8.0cm]{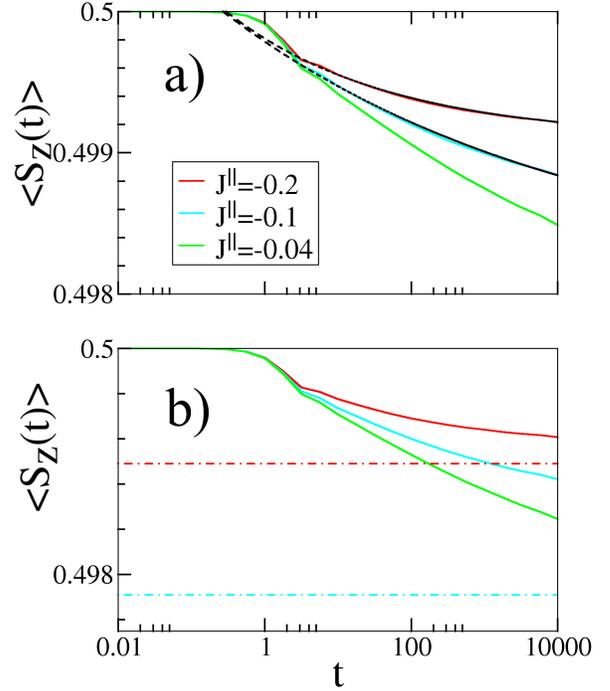}
\caption{\label{anisolong} Long-time result of the magnetization relaxation for anisotropic couplings and
comparison with a long-time fit. All parameters are equivalent to those from Fig.~\ref{anisoshort}.
In panel a), the full numerical solution of the flow equations (full lines in both panels)
is fitted against the analytical powerlaw behaviour
stated in Eq.~\eqref{anisoresult} (dashed lines). At time scales where renormalization of $J^\perp$ by $J^\parallel$
becomes significant and leads to a deviation from the perturbative result, again the asymptotic fit
describes the relaxation process very well. In panel b), a comparison with the saturation value
$\langle S^z(t \rightarrow \infty)\rangle=1/2+\alpha^2 \rho_F^2 /(4\tilde{j}^\parallel)$ (dashed lines) is given.
Clearly visible, anisotropic couplings $J^\perp < J^\parallel$ lead to much faster saturation of the
magnetization than isotropic couplings. Fitting parameters are given in Table \ref{table2}.}
\end{figure}

\begin{table}
\caption{\label{table2} For large times, we fitted the full numerical solutions
of the flow equations against the function $at^{-\sqrt{J_\parallel^2-J_{\perp}^2}}+b$,
as depicted in Fig.~\ref{anisolong}a . For couplings $-j^\parallel \leq 0.1$, quantitative agreement with the analytical result
$\langle S^z(t)\rangle=0.5\bigl(1 - \alpha^2/(2\tilde{j}^\parallel)t^{2\tilde{j}^\parallel}  +\alpha^2/(2\tilde{j}^\parallel)+\mathcal{O}(j^2)\bigr)$
is very good.}
\begin{ruledtabular}
\begin{tabular}{c| c c }
 $\boldsymbol{\langle S^z(t)\rangle= at^{-\sqrt{J_\parallel^2-J_{\perp}^2} }+b}$ & $j^\parallel=-0.05$ & $j^\parallel=-0.1$ \\
\hline
a & $1.66 \cdot 10^{-3}$ & $7.07 \cdot 10^{-4}$ \\
b & 0.4981        & 0.4991 \\
\end{tabular}
\end{ruledtabular}
\end{table}

We proceed as in the isotropic regime and plug the approximate couplings $\tilde{\gamma}_{\vec{k}\vec{k}^\prime}$
from (\ref{gammaaniso}) into the formal expression from  Eq.~(\ref{formalht})
to obtain the coefficient $h(t)$. This result is at least correct in the long-time limit.
In this way, we obtain the time-dependent magnetization as
\begin{equation}
\langle S^z (t)\rangle =\frac{1}{2}-\frac{1}{2}\int_0^{2D} \alpha^2 \frac{ k(x) }{x^{2+2\tilde{j}^\| }} ( 1 -\cos (xt) ) dx.
\label{resultaniso}
\end{equation}
As argued before, for anisotropies $|J^\parallel/J^\perp| \gtrsim 2$, it is possible to set
$\alpha \approx j^\perp$. In this case, again the short-time limit of the flow equation
result for $\langle S^z(t)\rangle$ coincides with the exact perturbative result of Eq. (\ref{shorttime}). Since our
approximations become only exact at low energies, we have in this case no
argument why they should in general hold in the short-time limit, where the influence of
higher energy scales might invalidate our approximations. In Fig.~\ref{anisoshort}, the
agreement of the numerical solution of the anisotropic flow equations with the perturbative short-time formula Eq.
(\ref{shorttimeresult}) is shown. Anisotropy due to increasing $|J^\parallel|$ does not
influence the short-time limit of the observable $\langle S^z(t) \rangle$, since $J^\parallel$ enters this quantity only in third order of
perturbation theory.

The long-time tail of the magnetization curve can be obtained as in the isotropic case by
replacing $\int_0^{\infty} dx\cos(xt) \dots$ with $\int_0^{f(t)/t}dx \dots~$, where
$f(t)$ is a function with values of $\mathcal{O}(1)$. This yields the asymptotic result
\begin{equation}
\langle S^z(t)\rangle = \frac{1}{2}
\left(1 - \frac{\alpha^2}{2\tilde{j}^\parallel}t^{2\tilde{j}^\parallel} + \frac{\alpha^2}{2\tilde{j}^\parallel}+O(j^2)\right).
\label{anisoresult}
\end{equation}
In Fig.~\ref{anisolong}, the qualitative difference between the anisotropic and 
isotropic couplings is depicted. Clearly visible, anisotropy leads to qualitatively faster saturation of the
magnetization to its steady state value. The algebraic long-time tail of the magnetization
curve sets in immediately after the perturbative short-time limit. A numerical fit of the
long-time tails yields very good agreement with the analytical prediction of Eq.
(\ref{anisoresult}), as demonstrated in table~\ref{table2}. Furthermore, our calculations show
that the steady state magnetization
$\langle S^z(t\rightarrow \infty) \rangle = \frac{\alpha^2}{2\tilde{g}_\parallel}$
is again reduced twice as much from full polarization than in equilibrium.

\subsection{Conclusions}

The numerical results show that spin relaxation of ferromagnetic impurities is described by
two different regimes, which are well described within the flow equation method. An initial short-time regime,
driven by unrenormalized spin-flip scattering with an amplitude  $\propto J^{\perp2}$ can be described
within unrenormalized perturbation theory. At low-energy scales (corresponding to large time scales),
renormalization of spin-flip scattering slows down the relaxation rate.
Quite naturally, isotropic couplings are renormalized only logarithmically in comparison to algebraic renormalization
of $J^\perp$ in the anisotropic case, leading to much faster relaxation of anisotropically coupled spins.
It is important to note that our results are to leading order in $J$ independent on the spin quantum number $S$ of the impurity spin.
As detailed in appendix~\ref{highS}, a generalization to arbitrary $S$ leads to
\begin{equation}
\langle S^z (t)\rangle= Sh(t) ,
\end{equation}
where $h(t)$ is the universal function which we calculated above.


\section{\label{summary} Summary and discussion}

We have discussed the real time evolution of the ferromagnetic Kondo model initially
prepared in the product state
\begin{equation}
|\psi\rangle = |\uparrow \rangle \otimes |FS\rangle,
\label{initialstate}
\end{equation}
where $|FS\rangle$ is the non-interacting Fermi sea. Using the flow-equation method,
we have shown that the time-dependent magnetization $\langle S^z(t)\rangle_{\psi}$ approaches
a nonvanishing asymptotic value $\langle S^z(t\rightarrow\infty)\rangle_{\psi}$ logarithmically
(with a power law) in time if the model is isotropic (anisotropic). One key observation
is the fact that $\langle S^z(t\rightarrow\infty)\rangle_{\psi}$ differs from the equilibrium
value $\langle S^z\rangle_{\rm eq}$ for infinitesimal positive magnetic field. {\em In particular,
this implies that the system retains a memory of its initial preparation for all times.}

While this might appear surprising at first sight, it just reflects the following well-known property
of the equilibrium model: If one prepares the system in the initial state (\ref{initialstate})
and then slowly switches on the ferromagnetic coupling to the leads, according to the adiabatic theorem
the system will evolve to the
equilibrium ground state with magnetization $\langle S^z\rangle_{\rm eq}$. However, if the initial
state were $|\downarrow \rangle \otimes |FS\rangle$, then the asymptotic magnetization would be inverted.
In this sense even the equilibrium model retains a memory of its initial preparation due to
ergodicity breaking.

In the weak-coupling limit we could prove that asymptotic non-equilibrium magnetization vs.\
equilibrium magnetization differ by a factor two:
\begin{equation}
\left(\langle S^z(t\rightarrow\infty)\rangle_{\psi}\:- \frac{1}{2}\right)= 2\,\left(\langle S^z\rangle_{\rm eq}\:-\frac{1}{2} \right)
\end{equation}
This factor two occurs in a general class of discrete systems \cite{Moeckel2008,MoeckelKehrein_AnnPhys}
in weak-coupling perturbation theory. In a simple exactly solvable toy model with two lead levels,
we could explicitly evaluate this factor $r(g)$ beyond its weak-coupling limit $r(g=0)=2$, see Fig.~\ref{toyratio}.
The toy model results for positive (that is antiferromagnetic) coupling~$g$ are of course not relevant for the
actual Kondo model due to the strong-coupling divergence that makes perturbation theory invalid for $g>0$.
On the other hand,
the fact that the running coupling flows to zero on the ferromagnetic side is just the
reason why our analytical flow-equation results become asymptotically exact for small negative~$g$.
This is also supported by exact numerical results using time-dependent NRG \cite{hackl_prl}
which showed excellent agreement with the analytical flow-equation calculation for weak ferromagnetic coupling.
One can even understand the differences between the exact numerical results and the perturbative
flow-equation results in Ref.~\onlinecite{hackl_prl}: According to Fig.~\ref{toyratio} one would expect the
flow-equation calculation to overestimate the non-equilibrium reduction of the magnetization (since $r(g)<2$ for $g<0$
in Fig.~\ref{toyratio}), which is exactly what one observes in Ref.~\onlinecite{hackl_prl} for larger ferromagnetic coupling.

As discussed in Ref.~\onlinecite{hackl_prl}, our results for the ferromagnetic Kondo model
have direct applications to molecular quantum dots. For instance, the magnetic dynamics of single-molecule
magnets coupled to metallic leads may be captured (in the co-tunneling regime) by a ferromagnetic Kondo
model.\cite{leuenberger} Another experimentally relevant situation are coupled semiconductor quantum dots
with partially screened spins.

On the methodological side, we have shown that the flow-equation method can be applied
successfully to real-time evolution problems in quantum impurity models.
We envision further applications of our method in the study of non-equilibrium dynamics
near impurity quantum phase transitions.\cite{mv_rev}


\begin{acknowledgments}
We acknowledge discussions with M. Garst, W. Hofstetter, D. Roosen, and A. Rosch.
This research was supported by the DFG through SFB~608, SFB-TR/12, and FG 960.
S.~K. also acknowledges support through the Center for Nanoscience
(CeNS) Munich and the German Excellence
Initiative via the Nanosystems Initiative Munich (NIM).
\end{acknowledgments}


\appendix

\section{\label{toyalgebra} Matrix representation of the toy model}

The five basis states with total spin $S=1/2$, $S^z=1/2$ and two fermions in
the electronic levels are given by
\begin{eqnarray}
| \psi_0 \rangle &=&| 0\rangle \otimes | \uparrow \downarrow \rangle \otimes | \uparrow \rangle\nonumber \\
| \psi_1 \rangle &=&| \uparrow \rangle \otimes | \uparrow \rangle \otimes | \downarrow \rangle \nonumber\\
| \psi_2 \rangle &=&| \downarrow \rangle \otimes | \uparrow \rangle \otimes | \uparrow \rangle \nonumber\\
| \psi_3 \rangle &=&| \uparrow \downarrow \rangle \otimes | 0 \rangle \otimes | \uparrow \rangle \nonumber\\
| \psi_4 \rangle &=&| \uparrow \rangle \otimes | \downarrow \rangle \otimes | \uparrow \rangle
\end{eqnarray}
where the first ket in the tensor product corresponds to the $c$-electron in our
toy model, the second ket corresponds to the $d$-electron and the third ket corresponds
to the impurity spin.
The subspace spanned by these five states is denoted by ${\mathcal H}_{2,\frac{1}{2}}$.
Due to the $SU(2)$-symmetric interaction, the time-evolved initial state
as well as the interacting ground state in the presence of an infinitesimal positive
magnetic field lie in ${\mathcal H}_{2,\frac{1}{2}}$ .

In the subspace ${\mathcal H}_{2,\frac{1}{2}}$, the Hamiltonian
(\ref{toy}) of the toy model is represented
by the matrix $H=H_0+g\,H_{int}$, $H_{i+1,j+1}\stackrel{\rm def}{=}\langle \psi_i|H|\psi_j \rangle$
for $i,j=0,1,2,3,4$:
\begin{equation}
H_0=\left(
\begin{array}{ccccc}
-2 & 0 & 0 & 0 & 0 \\
 0 & 2 & 0 & 0 & 0  \\
 0 & 0 & 0 & 0 & 0 \\
 0 & 0 & 0 & 0 & 0  \\
 0 & 0 & 0 & 0 & 0 \\
\end{array}
\right)
\end{equation}
\begin{equation}
H_{int}=\left(
\begin{array}{ccccc}
 0 & \frac{1}{2} & \frac{1}{2} &0&0 \\
 \frac{1}{2} & 0 & 0 & 0 & 0  \\
 \frac{1}{2} & 0 & -\frac{1}{4} & 0 & 0 \\
 0 & 0 & 0 & 0 & 0  \\
 0 & 0 & 0 & 0 & 0 \\
\end{array}
\right) \ .
\label{5bands}
\end{equation}
The states $|\psi_3\rangle$ and $|\psi_4\rangle$ decouple, so that one effectively
only has to diagonalize a $3\times 3$-matrix to i) solve the dynamics of the
toy model exactly and to ii) determine the ground state magnetization.
Since the diagonalization is a trivial step with lengthy expressions,
we will not give details here.

However, it is interesting to see explicitly how the factor~2 in the ratio $r(g)$
from (\ref{defrg}) comes about in the weak-coupling limit. In perturbation theory
the interacting eigenstates are given by
\begin{equation}
|\tilde\psi_j\rangle=|\psi_j\rangle + g\,\sum_{i\neq j}
|\psi_i\rangle\,\frac{\langle \psi_i|H_{int}| \psi_j\rangle}{E_j-E_i}+\mathcal{O}(g^2) \ ,
\end{equation}
where $E_0=-2, E_1=2, E_2=0$ are just the eigenvalues of $H_0$.
For the observable $\hat{O}=S^z-1/2$ one has
$\langle \psi_0|\hat{O}|\psi_0 \rangle =  \langle \psi_1|\hat{O}|\psi_1 \rangle=0$ and
therefore the ground state expectation value
\begin{eqnarray*}
O_{eq}&=&\langle \tilde\psi_0|\hat{O}|\tilde\psi_0 \rangle \nonumber \\
&=&g^2\,\langle \psi_2|\hat{O}|\psi_2\rangle\:\frac{|\langle \psi_2|H_{int}|\psi_0\rangle|^2}{(E_2-E_0)^2}
+\mathcal{O}(g^3) \ .
\end{eqnarray*}
The time-averaged expectation value starting from the initial state $|\psi_0\rangle$
is according to (\ref{mean})
\begin{equation}
\overline{O(t)}=\sum_j |\langle \tilde\psi_j|\psi_0\rangle|^2\:
\langle \tilde\psi_j|\hat{O}|\tilde\psi_j\rangle
\end{equation}
and one can easily verify that only the terms $j=0,2$ contribute plus corrections
in order~$g^3$. Each of these terms is identical to $O_{eq}$ plus again corrections
in order~$g^3$, which proves
\begin{equation}
r(g)=\frac{\overline{O(t)}}{O_{eq}}=2+\mathcal{O}(g) \ .
\end{equation}
The general proof in Ref.~\onlinecite{MoeckelKehrein_AnnPhys} is a generalization of this argument.
The universal factor~2 in the weak-coupling limit plays e.g. a key role
in the thermalization of a Fermi liquid after an interaction quench, that
is the opposite limit of the adiabatic Landau Fermi-liquid
paradigm.\cite{Moeckel2008,MoeckelKehrein_AnnPhys}


\section{\label{highS} Flow equations for general spin S}

We now analyze how the magnetization curve $\langle S^z (t) \rangle$ is modified
if the impurity spin has a general quantum number $S > \frac{1}{2}$.
The commutation relation $[S^i,S^j]_{-}=i \epsilon_{ijk} S^k$ is independent
of $S$, and we will show that also the flow equations for
the spin components $S^a$ remain unchanged to leading order
in $J$. We illustrate the proof only for the case of
isotropic couplings, since the prove for
anisotropic couplings requires only slight modifications.

Using the ansatz (\ref{szb}) together with the generator (\ref{generator}),
it is readily seen that modifications to the flow equations for $S^a(B)$
can only arise due to the commutator
\begin{equation}
[:\vec{S} \cdot \vec{s}_{\vec{t}\vec{t^\prime}}:,(:\vec{S} \times \vec{s}_{\vec{u^\prime} \vec{u}}:)^a]_- \ .
\label{commute}
\end{equation}
We now decompose
\begin{equation}
S^i S^j=x_{ij}+\frac{i}{2} \sum_{k} \epsilon_{ijk}S^k \ ,
\label{decomposespin}
\end{equation}
where $\frac{i}{2}\sum_k \epsilon_{ijk} S^k$ is the operator obtained by
projecting  $S^i S^j$ on the operator $\sum_k \epsilon_{ijk} S^k$ as a basis
operator in spin Hilbert space and $x_{ij}$ is the operator
defined by Eq.~\eqref{decomposespin}.
Using this decomposition in Eq. (\ref{commute}), we obtain
\begin{eqnarray}
&&[:\vec{S} \cdot \vec{s}_{\vec{t}\vec{t^\prime}}:,(:\vec{S} \times \vec{s}_{\vec{u^\prime} \vec{u}}:)^a]_- = \nonumber\\
&&\sum_{ijk\alpha\beta\mu \nu } \biggl( \frac{1}{4} x_{ij}
[\sigma_{\alpha \beta}^i \sigma_{\mu \nu}^k \epsilon_{ajk}[:c_{\vec{t^\prime} \alpha}^\dagger c_{\vec{t}\beta}^\dagger:,:c_{\vec{u^\prime}\mu}^\dagger c_{\vec{u} \nu}^\pdag:]_-] \nonumber\\
&&+ \frac{1}{4} \frac{i}{2} \sum_{k^\prime} \sigma_{\alpha \beta}^i \sigma_{\mu \nu}^k \epsilon_{ajk} \epsilon_{ijk^\prime} S^{k^\prime}
[:c_{\vec{t^\prime} \alpha}^\dagger c_{\vec{t}\beta}^\dagger:,:c_{\vec{u^\prime}\mu}^\dagger c_{\vec{u} \nu}^\pdag:]_+ \biggr) \ .\nonumber\\
\label{rewritten}
\end{eqnarray}
It is now readily seen that corrections to the flow equations Eq.~\eqref{feqsz} can only arise from the term in
Eq.~\eqref{rewritten} that contains the operator $x_{ij}$.
This term can be rewritten as
\begin{equation}
\frac{i}{8}\sum_{ijkl} x_{ij} \epsilon_{ajk} \epsilon_{lik}[ s_{\vec{t^\prime} \vec{u}}^l \delta_{\vec{t} \vec{u^\prime}} + \delta_{\vec{t^\prime} \vec{u}} s_{\vec{u^\prime} \vec{t}}^l] \ ,
\end{equation}
and due to the decomposition assumption~\eqref{decomposespin}, it is readily seen that this
expression has a vanishing projection on the operator $(:\vec{S} \times \vec{s}_{\vec{u^\prime} \vec{u}}:)^a$.
Thus it can only be generated as a subleading correction in $\mathcal{O}(J^2)$ of the flowing spin operators.
From the identity $\sum_{i} x_{ii}=S(S+1)$,
it is finally seen that the flowing spin components $S^a$ have corrections which are of $\mathcal{O}(J^2S^2)$.
For the magnetization curve $\langle S^z(t)\rangle$, this means that the relation
\begin{equation}
\langle S^z(t)\rangle = 2S\langle S^z(t) \rangle|_{S=\frac{1}{2}} +\mathcal{O}(J^2S^2)
\end{equation}
is fulfilled.


\section{\label{tree} Validity of the tree-level approximation}

In section V, we made use of the approximative differential equation
\begin{equation}
\frac{d\gamma_{\vec{k^\prime} \vec{k}}}{dB} = h(B)(\varepsilon_{\vec{k^\prime}} -
\varepsilon_{\vec{k}}) J_{\vec{k^\prime} \vec{k}}^\perp(B) +\mathcal{O}((\varepsilon_{\vec{k^\prime}} - \varepsilon_{\vec{k}})J^2) \ .
\label{treeflowapp}
\end{equation}
In case of anisotropic couplings $|J^\perp|<|J^\parallel|$ and $J^\parallel<0$,
the couplings $J_{\vec{k}^\prime \vec{k}}^\parallel$ contained in the neglected terms flow to a finite value at the
Fermi surface (given by $\epsilon_{\vec{k}}=\epsilon_{\vec{k}^\prime}$=0),
while $J_{\vec{k^\prime} \vec{k}}^\perp(B)$ will flow to zero.
A justification of neglecting all terms of $\mathcal{O}(J^2)$ in $\frac{d\gamma_{\vec{k^\prime} \vec{k}}}{dB}$
has therefore to consider the flow equations including all terms of up to $\mathcal{O}(J^2)$,
\begin{eqnarray}
\frac{d\gamma_{\vec{k}^\prime \vec{k}} }{dB} &=& h(B) \Delta \epsilon \frac{J_{IR}^\perp (B)}{N}e^{-B(\Delta \epsilon)^2} \nonumber\\
&-& \frac{1}{4}\int_{-D}^D d\epsilon \, \text{sgn}(\epsilon) \biggl[ (\Delta \epsilon +\epsilon_{\vec{k}^\prime} -\epsilon)\frac{J_{IR}^\parallel (B)}{N} \nonumber\\
&\times& e^{-B(\Delta\epsilon+\epsilon_{\vec{k}^\prime} -\epsilon)^2} \int_0^{B} dB^\prime (\epsilon -\epsilon_{\vec{k}^\prime})J_{IR}^\perp (B^\prime) \nonumber\\
&\times& e^{-B^\prime(\epsilon -\epsilon_{\vec{k}^\prime})^2} +(\epsilon_{\vec{k}^\prime}-\epsilon)\frac{J_{IR}^\parallel(B)}{N}e^{-B(\epsilon-\epsilon_{\vec{k}^\prime})^2}\nonumber\\
&\times&\int_0^B dB^\prime(\Delta \epsilon +\epsilon_{\vec{k}^\prime} -\epsilon) \nonumber\\
&\times&J_{IR}^\perp (B^\prime) e^{-B^\prime (\epsilon-\Delta\epsilon-\epsilon_{\vec{k}^\prime} )^2}\biggr] \ ,
\label{derivative1}
\end{eqnarray}
where we defined $\Delta \epsilon = \epsilon_{\vec{k}}-\epsilon_{\vec{k}^\prime}$
and made use of the infrared parametrization~\eqref{infrared}.
Only in three integration intervalls of width $\Delta \epsilon$, the integrand in Eq.~\eqref{derivative1} is nonzero
and we can simplify $\frac{d\gamma_{\vec{k} \vec{k}^\prime} }{dB}$ according to
\begin{eqnarray}
\frac{d\gamma_{\vec{k}^\prime \vec{k}} }{dB} &=& \Delta \epsilon  \frac{J_{IR}^\perp (B)}{N} e^{-B(\Delta \epsilon)^2} \nonumber\\
&+&\frac{1}{2}\int_{-\Delta \epsilon}^0 d\epsilon \bigl[ (\Delta \epsilon +\epsilon_{\vec{k}^\prime}-\epsilon)\frac{J_{IR}^\parallel(B)}{N}
e^{-B(\Delta\epsilon +\epsilon_{\vec{k}^\prime} -\epsilon)^2} \nonumber\\
&\times& \int_0^B dB^\prime (\epsilon-\epsilon_{\vec{k}^\prime}) J_{IR}^{\perp}(B^\prime)e^{-B^\prime(\epsilon-\epsilon_{\vec{k}^\prime})^2}\bigr] \nonumber\\
&+&\mathcal{O}\bigr(e^{-BD^2}\bigl)  \ ,
\label{evaluatedder}
\end{eqnarray}
where $\mathcal{O}\bigr(e^{-BD^2}\bigl)$ arises from integrals close to the band edges.
Since the perpendicular coupling $J_{IR}^\perp(B)$ asymptotically decays as
$J_{IR}^\perp(B)\propto\frac{\alpha}{\rho_F}B^{\tilde{j}^\parallel/2}$ (see sec.~\ref{anisosec}),
a lower boundary for the contribution of $\mathcal{O}(J)$ in~\eqref{evaluatedder} is
\begin{equation}
\Delta \epsilon  \frac{\alpha}{\rho_F}B^{\tilde{j}^\parallel/2} e^{-B(\Delta \epsilon)^2} \ ,
\end{equation}
while it is readily seen that the contribution of $\mathcal{O}(J^2)$ has the upper boundary
\begin{equation}
BJ^\perp J^\parallel \mathcal{O}(\Delta \epsilon)^3
\end{equation}
as long as $B \lesssim (\Delta \epsilon)^{-2}$.
Hence, in the regime $B \lesssim (\Delta \epsilon)^{-2}$ we can approximate
$\frac{d\gamma_{\vec{k}^\prime \vec{k}} }{dB} \approx \Delta \epsilon  J^\perp (B) e^{-B(\Delta \epsilon)^2}$ if
\begin{equation}
\Delta \epsilon^2 B^{1-\tilde{j}^\parallel/2}  \ll \frac{1}{j^\parallel} \ .
\end{equation}
The remaining regime $B \gtrsim (\Delta \epsilon)^{-2}$ is described by
exponential decay of $\frac{d\gamma_{\vec{k}^\prime \vec{k}} }{dB}$
and is not of importance. In conclusion, we showed that the
approximation
\begin{equation}
\frac{d\gamma_{\vec{k^\prime} \vec{k}}}{dB} \approx (\varepsilon_{\vec{k^\prime}} -
\varepsilon_{\vec{k}}) J_{\vec{k^\prime} \vec{k}}^\perp(B)
\end{equation}
is valid also for anisotropic couplings with $|J^\perp|<|J^\parallel|$ and $J^\parallel<0$.


\section{\label{alternative} An alternative route of calculation for Eq. (\ref{magnetization})}

An alternative derivation of the formal result~\eqref{formalht} for the magnetization
$\langle S^z(t)\rangle=h(t)/2$ exploits the sum rule~\eqref{sumrule}.
Since the operators $S^z(t)$ and $\tilde{S}^z(t)$ are related to the flowing operator $S^z(B)$
by a unitary transformation, the sum rule $\langle S^z(B)^2\rangle=1/4$ applies for both operators,
and by subtracting the corresponding sum rule for expression (\ref{tildeszt}) from that of expression
(\ref{szt}), the coefficient $h(t)$ can be expressed as
\begin{eqnarray}
h^2(t) &=& \tilde{h}^2 + \sum_{\vec{k}\vec{k^\prime}} (| \tilde{\gamma}_{\vec{k^\prime} \vec{k}} (t)|^2-| \gamma_{\vec{k^\prime} \vec{k}} (t) |^2 )\nonumber\\
&\times& n(\vec{k^\prime} ) (1-n(\vec{k})) + \mathcal{O}(J^2) \ .
\label{sumrules}
\end{eqnarray}
Eq. (\ref{sumrules}) is another starting point to calculate the magnetization
$h(t)/2$. The coefficients $\tilde{\gamma}_{\vec{k}^\prime \vec{k}} (t)$ and $\gamma_{\vec{k^\prime} \vec{k}} (t)$
in~\eqref{sumrules} are given by Eqs~\eqref{tildeszt} and~\eqref{treelevel}, and we obtain
\begin{eqnarray}
h^2(t)&=&\tilde{h}^2+\sum_{\vec{k}\vec{k^\prime}}|\tilde{\gamma}_{\vec{k}\vec{k^\prime}}|^2\Bigl(-1+2\cos[(\varepsilon_{\vec{k}} -\varepsilon_{\vec{k^\prime}})t]\Bigr)\nonumber\\
&\times& n(\vec{k^\prime} ) (1-n(\vec{k})) + \mathcal{O}(J^2) \ .
\label{sumrules2}
\end{eqnarray}
Using the known result $\tilde{h}=1+\mathcal{O}(J)$, it is easily seen that
the right hand side of Eq. (\ref{sumrules2}) behaves as $1+\mathcal{O}(J)$,
and the expansion of $\sqrt{h^2(t)}$ to first order in the coupling $J$ is therefore
\begin{eqnarray}
h(t)&=&\tilde{h}+\sum_{\vec{k}\vec{k^\prime}} |\tilde{\gamma}_{\vec{k}\vec{k^\prime}}|^2 \Bigl( \cos[(\varepsilon_{\vec{k}} -\varepsilon_{\vec{k^\prime}})t]-\frac{1}{2}\Bigr)\nonumber\\
&\times& n(\vec{k^\prime} ) (1-n(\vec{k})) + \mathcal{O}(J^2) \ .
\end{eqnarray}
This result is equivalent to Eq.~\eqref{formalht}.


\section{\label{diagonal} Diagonal parameterization of isotropic couplings}

In section~\ref{analytical}, we integrated the flow equation
\begin{equation}
\frac{d\gamma_{\vec{k}\vec{k}^\prime}}{dB}=(\epsilon_{\vec{k}}-\epsilon_{\vec{k}^\prime})J_{\vec{k}\vec{k}^\prime}^\perp(B)
\label{diaggammaflow}
\end{equation}
by using the approximative parameterization $J_{\vec{k}\vec{k}^\prime}^\perp(B)=(J_{IR}(B)/N)e^{-B(\epsilon_{\vec{k}}-\epsilon_{\vec{k}^\prime})^2}$.
This parameterization assumes isotropic couplings $J_{\vec{k}\vec{k}^\prime}^\perp(B) \equiv J_{\vec{k}\vec{k}^\prime}^\parallel(B)$
and requires low energies $|\epsilon_{\vec{k}}|\ll D\,,|\epsilon_{\vec{k}^\prime}|\ll D$.
Here, we provide a more general integration of Eq.~\eqref{diaggammaflow}
by parameterizing the flow of the couplings $J_{\vec{k}\vec{k^\prime}}(B)$ by an ansatz that
also describes couplings above the low-energy limit appropriately.
One makes the ansatz
\begin{equation}
J_{ \vec{k}\vec{k^\prime}}(B)= J_{\overline{ \vec{k}\vec{k}^\prime}}(B) e^{-B(\varepsilon_{\vec{k}} - \varepsilon_{\vec{k}^\prime} )^2},
\label{diagonalparam}
\end{equation}
with $J_{\overline{ \vec{k}\vec{k^\prime}}}(B=0)= J/N <0$, where
$\overline{\vec{k^\prime} \vec{k}}$ is a label for the
energy median
\begin{equation}
\varepsilon_{\overline{\vec{k}^\prime\vec{k}}}\stackrel{\rm def}{=}\frac{\varepsilon_{\vec{k}}+\varepsilon_{\vec{k^\prime}}}{2} \ ,
\end{equation}
such that the couplings $J_{\overline{\vec{k^\prime} \vec{k}}}(B)$ depend only on the
energy $\varepsilon_{\overline{\vec{k}\vec{k}^\prime}}$ and $B$.
This approach has been dubbed {\sl diagonal parameterization}, since it approximates
the full set of $N \times N$ couplings by its $N$ diagonal entries. It has been used previously
in the context of the non-equilibrium Kondo model in Ref.~\onlinecite{KehreinKMV}, and a
detailed discussion is given in Ref.~\onlinecite{Kehrein_STMP}. It is trivial that this
ansatz becomes asymptotically exact for the low-energy couplings near the Fermi surface,
since the coupling $J_{\overline{\vec{k} \vec{k}^\prime}}(B)$ will reduce to the IR-parameterization $J_{IR}(B)$
used in sec.~\ref{flowequation} once $\epsilon_{\vec{k}} , \epsilon_{\vec{k}^\prime} \rightarrow 0$.
In addition, it describes the deviations from the flow of $J_{IR}(B)$ caused by interactions between matrix elements of higher energy
scales.

The number of flow equations for the couplings reduced now from $N \times N$ to the $N$ differential equations
(using $J_{\vec{k}}\stackrel{\rm def}{=}J_{\overline{\vec{k} \vec{k}}}$)
\begin{eqnarray}
\frac{d J_{\vec{k}}(B)}{dB}
&=& \sum_{\mu} 2(\varepsilon_{\vec{k}} + \varepsilon_{\vec{\mu}})
J_{\overline{\vec{k} \vec{\mu}}} J_{\overline{ \vec{\mu} \vec{k}}} e^{-2B(\epsilon_{\vec{k}}-\epsilon_{\vec{\mu}})^2}
[n(\vec{\mu} ) - \frac 1 2] \nonumber\\
&+&\mathcal{O}(J^2) \ .
\label{diagonalflow}
\end{eqnarray}
Off-diagonal couplings with huge energy differences decay exponentially fast, and it is
possible to make the approximation $J_{\overline{\vec{k}\vec{k^\prime}} }(B) \simeq
J_{\vec{k}}(B)$ in Eq.~(\ref{diagonalflow}). Furthermore, the initial condition for
$J_{\vec{k}}(B)$ can be adjusted to $J_{\vec{k}}(B=D^{-2})=\frac{J}{N}$, since the flow
of $J_{\vec{k}}(B)$ is unrenormalized for energies near the band cut-off, corresponding
to $B \lesssim D^{-2}$. Now, the diagonal couplings flow approximately as
\begin{eqnarray}
\frac{d J_{\vec{k}}(B)}{dB} &=& \sum_{\vec{\mu}} 2(\varepsilon_{\vec{k}} - \varepsilon_{\vec{\mu}})e^{-2B(\epsilon_{\vec{k}}-\epsilon_{\vec{\mu}})^2}
J_{\overline{\vec{k} \vec{\mu}}} J_{\overline{ \vec{\mu} \vec{k}}} (n(\vec{\mu} ) -\frac{1}{2}) \nonumber\\
&\simeq& -\frac{J_{\vec{k}}^2(B) \rho_F}{2 BN} e^{-2B \varepsilon_{\vec{k}}^2} \ ,
\end{eqnarray}
where we evaluated the momentum sum at $T=0$.
This differential equation can be integrated in the closed form
\begin{equation}
J_{\vec{k}}(B) = \frac{J}{N}
\left(1 - J \frac{\rho_F}{2} \int_{D^{-2}}^B dB^\prime \frac{\exp(-2B^\prime\varepsilon_{\vec{k}}^2)}{B^\prime}\right)^{-1} \ .
\end{equation}
In order to evaluate the couplings $\tilde{\gamma}_{\vec{k}\vec{k^\prime}}$
by integrating Eq.~\eqref{diaggammaflow}, we approximate the couplings $J_{\vec{k}}(B)$ as
$J_{\vec{k}}(B=(\varepsilon_{\vec{k}}-\varepsilon_{\vec{k^\prime}})^{-2})$. This
is justified up to corrections of $\mathcal{O}(J^2)$, since $J_{\vec{k}}(B)$ depends
logarithmically or weaker on the parameter $B$. We can further
evaluate the couplings $J_{\vec{k}}(B=(\varepsilon_{\vec{k}}-\varepsilon_{\vec{k^\prime}})^{-2})$
by performing the sequence of approximations
\begin{eqnarray}
&&\int_{D^{-2}}^{(\varepsilon_{\vec{k}}-\varepsilon_{\vec{k^\prime}})^{-2}} dB^\prime
\frac{\exp(-2B^\prime \varepsilon_{\vec{k}}^2)}{B^\prime} \nonumber\\
&&=\int_{D^{-2}}^{(\varepsilon_{\vec{k}}-\varepsilon_{\vec{k^\prime}})^{-2}} dB^\prime
\frac{\exp(-2B^\prime (\varepsilon_{\vec{k}}-\varepsilon_{\vec{k^\prime}})^{2})}{B^\prime}+\mathcal{O}(1) \nonumber\\
&&=\int_{D^{-2}}^\infty dB^\prime
\frac{\exp(-2B^\prime (\varepsilon_{\vec{k}}-\varepsilon_{\vec{k^\prime}})^{-2})}{B^\prime}+\mathcal{O}(1) \nonumber\\
&&=-\gamma - \ln \Bigl( \frac{2(\varepsilon_{\vec{k}}-\varepsilon_{\vec{k^\prime}})^2}{D^{2}}\Bigr)
 +\mathcal{O}\Bigl( \frac{2(\varepsilon_{\vec{k}} -\varepsilon_{\vec{k^\prime}})^2 }{D^2} \Bigr) +\mathcal{O}(1) \ . \nonumber\\
\end{eqnarray}
We assumed in the second line that $\epsilon_{\vec{k}}$ and $\epsilon_{\vec{k}}-\epsilon_{\vec{k}^\prime}$ are of the same
order, since the phase space where this is not the case has subleading measure in momentum sums $\sum_{\vec{k}\vec{k}^\prime}$.
Assuming furthermore $\bigl| \ln \Bigl( \frac{(\varepsilon_{\vec{k}}-\varepsilon_{\vec{k^\prime}})^2}{D^{2}}\Bigr)\bigr| \gg 1$
we made use of the asymptotic expansion of the exponential integral
\begin{equation}
Ei(1,x)=\int_{x}^\infty \frac{e^t}{t}dt=-\gamma - \ln x +\mathcal{O}(x)\nonumber\\
\end{equation}
in the last line, with $\gamma \approx 0.5772$.

The parameterization \eqref{diagonalparam} finally justifies that the coefficients
$\tilde{\gamma}_{\vec{k}\vec{k^\prime}}$ can be obtained using the
infrared parameterization $J_{\vec{k}\vec{k}^\prime}^\perp(B)=(J_{IR}(B)/N)e^{-B(\epsilon_{\vec{k}}-\epsilon_{\vec{k}^\prime})^2}$
in the limit $\bigl| \ln \Bigl( \frac{(\varepsilon_{\vec{k}}-\varepsilon_{\vec{k^\prime}})^2}{D^{2}}\Bigr)\bigr| \gg 1$.
This result is of particular importance for the the time dependence of the magnetization $\langle S^z(t)\rangle$, which is fully
described by the couplings  $\tilde{\gamma}_{\vec{k}\vec{k^\prime}}$ (see Eq. \eqref{formalht}).


\section{Normal ordering}

Within the flow equation approach, 
fermionic operators occuring in transformed observables are usually normal-ordered
in order to separate the various interaction terms generated during the
flow into irreducible objects in the spirit of a diagrammatic expansion using Wick's theorem.\cite{Kehrein_STMP} 
In practice this procedure is implemented with respect to a given reference state or density matrix, such 
that normal ordering of a product $c_{\vec{k}\sigma}^\dagger c_{\vec{k}^\prime \sigma^\prime}^\pdag$ 
of fermionic operators amounts to subtracting the expectation value with respect to the given reference state or density matrix,
\begin{equation}
:c_{\vec{k}\sigma}^\dagger c_{\vec{k}^\prime \sigma^\prime}:=c_{\vec{k}\sigma}^\dagger c_{\vec{k}^\prime \sigma^\prime}^\pdag
-\langle c_{\vec{k}\sigma}^\dagger c_{\vec{k}^\prime \sigma^\prime}^\pdag\rangle \ .
\end{equation}
In our particular problem, we used the initial state $| \psi_i \rangle$ as reference state
for the normal ordering procedure.
However, this reference state will look very different at some finite time $t$ or
if it is transformed into the representation of the flowing operator by the unitary transformation 
$U(B)$ induced by the flow equation approach.
In order to use a normal-ordering prescription with respect to 
the same quantum state for all values of $B$, we need to define 
normal ordering with respect to the time-evolved initial state in the basis for a
given value of $B$,
\begin{equation}
| \psi (B,t)\rangle \stackrel{\rm def}{=} U(B)e^{iHt} | \psi_i\rangle \ .
\end{equation}
The contraction of fermionic operators then depends both on 
$B$ and on time $t$,
\begin{equation}
n_{ll^\prime}(B,t)\stackrel{\rm def}{=}\langle  \psi_i | e^{-iHt} U^\dagger(B)  c_{l}^\dagger c_{l^\prime}^\pdag U(B) e^{iHt} | \psi_i\rangle \ ,
\label{fullcontractions}
\end{equation}
where we introduced a general multiindex $l$ comprising both spin and momentum label.
Since a calculation of the full dependence of $n_{ll^\prime}(B,t)$ on the parameters $B$ and $t$ 
turns out to be a very complicated problem of its own, we show here that the improvement
achieved by including this dependence in the  contractions $n_{ll^\prime}(B=0,t=0)$ 
enters the flow equations for spin operators (the Hamiltonian) only to $\mathcal{O}(J^2)$ (to $\mathcal{O}(J^3)$ ),
having no influence on the leading order of our calculation, which is
$\mathcal{O}(J)$ ($\mathcal{O}(J^2)$).

In order to reduce~\eqref{fullcontractions} to the form $n_{ll^\prime}(B,t)=n_{ll^\prime}(B=0,t=0)+\mathcal{O}(J)$,
we write the transformed operators $c_l^\pdag(B)=U^\dagger(B) c_l U(B)$ in the general form
\begin{equation}
c_l^\pdag(B)=f_l(B)c_l^\pdag + J \times \text{composite operator} + \mathcal{O}(J^2) \ ,
\end{equation}
with a flowing coupling parameter $f_l(B)$ and the unitary transfomation $U(B)$ generated by the generator \eqref{generator}.
Here and in the following we exploit that $f_l(B)$ behaves as $1+\mathcal{O}(J)$, since
$f_l^2(B)$ determines the quasiparticle weight of the conduction electrons that is
renormalized only perturbatively in $J$. 
In this way the contractions $n_{ll^\prime}(B,t)$ simplify to 
\begin{equation}
n_{ll^\prime}(B,t)=\langle  \psi_i | e^{iHt}   c_{l}^\dagger c_{l^\prime}^\pdag e^{-iHt} | \psi_i\rangle + \mathcal{O}(J)\ .
\end{equation}
The time-evolved fermionic operator  $e^{iHt} c_{l}^\pdag e^{-iHt} $ can now be formally 
calculated in the forward-backwards transformation scheme depicted in Fig., generating an effective
solution of the Heisenberg equation of motion for $c_l^\pdag$ of the form 
$c_l^\pdag(t)= d_l(t) c_l^\pdag + J \times \text{composite operator} + \mathcal{O}(J^2)$,
with a coupling function $d_l(t)=1+\mathcal{O}(J)$.
Now the contractions can be written as
\begin{equation}
n_{ll^\prime}(B,t) \langle  \psi (B,t) |   c_{l\sigma}^\dagger c_{l^\prime \sigma^\prime}^\pdag  | \psi (B,t)\rangle = \langle \psi_i| c_{l}^\dagger c_{l^\prime}^\pdag  | \psi_i \rangle +\mathcal{O}(J)\ .
\label{approxcontract}
\end{equation}
Since $n_{ll^\prime}(B,t)$ enters our flow equations only in terms of highest considered order in $J$
the correction of $\mathcal{O}(J)$ occuring in~\eqref{approxcontract} enters our calculation only to
subleading power in $J$. Therefore, it is sufficient in our calculation to use the simplified
contractions $n_{ll^\prime}=\langle \psi_i| c_{l}^\dagger c_{l^\prime}^\pdag  | \psi_i \rangle$
for normal ordering.



\begin{thebibliography}{11}
\bibitem{unitnote}
Here and in the following,
we employ units where $\hbar=k_B=1$.
\bibitem{footnotefullU}
The full unitary transformation $U$ can be expressed as an $B$-ordered
exponential, $U=T_{B}\exp\left(\int_{0}^{\infty}\eta(B)\,dB\right)$.
However, this expression is only formally useful since it cannot
be evaluated without additional approximations.
\bibitem{foot1}
In all numerical calculations, we will use a flat conduction electron band with energy range $[-D,D]$ and
the half width $D$ is chosen as $D=1$.
\end{thebibliography}
\end{document}